\documentclass[twocolumn]{aastex631}
\usepackage{amsmath}
\usepackage{CJK}
\usepackage{icomma}

\begin{document}
\begin{CJK*}{UTF8}{gbsn}
\title{Origin of the IRAS Vela Shell: New Insights from 3D Dust Mapping}

\correspondingauthor{Bore Gao}
\email{bgao6@jhu.edu}

\author[0000-0002-9902-6803]{Bore Annie Gao (高般若)}
\affiliation{Department of Physics and Astronomy\char`,{} Johns Hopkins University\char`,{} 3400 N. Charles Street\char`,{} Baltimore\char`,{} MD 21218\char`,{} USA}
\affiliation{Center for Astrophysics $|$ Harvard \& Smithsonian, 60 Garden St., Cambridge, MA 02138, USA}

\author[0000-0002-2250-730X]{Catherine Zucker}
\affiliation{Center for Astrophysics $|$ Harvard \& Smithsonian, 60 Garden St., Cambridge, MA 02138, USA}

\author[0000-0001-5921-5784]{Tirupati Kumara Sridharan}
\affiliation{National Radio Astronomy Observatory, Charlottesville, VA}

\author[0000-0001-9201-5995]{Cameren Swiggum}
\affiliation{University of Vienna, Department of Astrophysics, T\"{u}rkenschanzstraße 17, A-1180 Vienna, Austria}

\author[0000-0002-0404-003X]{Shmuel Bialy}
\affiliation{Technion-Israel Institute of Technology, Haifa, Israel}

\author[0000-0003-4852-6485]{Theo J. O'Neill}
\affiliation{Center for Astrophysics $|$ Harvard \& Smithsonian, 60 Garden St., Cambridge, MA 02138, USA}

\author[0000-0003-4797-7030]{J. E. G. Peek}
\affiliation{Space Telescope Science Institute, 3700 San Martin Drive, Baltimore, MD 21218, USA}

\author[0000-0001-7746-5461]{Luciana Bianchi}
\affiliation{Department of Physics and Astronomy\char`,{} Johns Hopkins University\char`,{} 3400 N. Charles Street\char`,{} Baltimore\char`,{} MD 21218\char`,{} USA}

\author[0000-0002-8109-2642]{Robert Benjamin}
\affiliation{University of Wisconsin,Whitewater, Whitewater, WI}

\author[0009-0007-1181-8034]{Lewis McCallum }
\affiliation{School of Physics and Astronomy, University of St Andrews, North Haugh, St Andrews, KY16 9SS, UK}

\author[0000-0003-1312-0477]{Alyssa Goodman}
\affiliation{Center for Astrophysics $|$ Harvard \& Smithsonian, 60 Garden St., Cambridge, MA 02138, USA}

\author[0000-0002-4355-0921]{Jo\~ao Alves}
\affiliation{University of Vienna, Department of Astrophysics, T\"{u}rkenschanzstraße 17, A-1180 Vienna, Austria}

\author[0000-0002-4658-7017]{Charles Lada}
\affiliation{Center for Astrophysics $|$ Harvard \& Smithsonian, 60 Garden St., Cambridge, MA 02138, USA}

\author[0000-0003-3122-4894]{Gordian Edenhofer}
\affiliation{Max Planck Institute for Astrophysics, Karl-SchwarzschildStraße 1, 85748 Garching bei M\"{u}nchen, Germany}
\affiliation{Ludwig Maximilian University of Munich, Geschwister-Scholl-Platz 1, 80539 M\"{u}nchen, Germany}

\author[0000-0002-0820-1814]{Rowan Smith}
\affiliation{School of Physics and Astronomy, University of St Andrews, North Haugh, St Andrews, KY16 9SS, UK}

\author[0000-0002-7365-5791]{Elizabeth Watkins}
\affiliation{Jodrell Bank Centre for Astrophysics, Department of Physics and Astronomy, University of Manchester, Oxford Road, Manchester M13 9PL, UK}

\author[0000-0001-7310-3375]{Kenneth Wood}
\affiliation{School of Physics and Astronomy, University of St Andrews, North Haugh, St Andrews, KY16 9SS, UK}

\author{Doni Anderson}
\affiliation{Department of Earth and Environmental Sciences, University of Michigan, 1100 N University Ave, Ann Arbor, MI 48109, USA}

\begin{abstract}

The IRAS Vela Shell (IVS) is a structure of enhanced FIR emission located towards the Gum Nebula, a prominent region of $\rm H\alpha$ emission in the local Milky Way shaped by various galactic stellar feedback over the past several million years. We constrain the 3D spatial geometry of the IVS using a parsec-resolution 3D dust map and contextualize it within the broader Gum Nebula. Our analysis reveals a dense, bowl-like IVS structure below the Galactic plane, with a more diffuse component above. We obtain a total shell mass of $5.1_{-2.4}^{+2.4}\times 10^{4}\;\rm M_{\odot}$ and, incorporating previous studies on shell expansion, a momentum of $6.0_{-3.4}^{+4.7}\times 10^{5}\;\rm M_{\odot}\;km\; s^{-1}$. We find a spatial correlation between the morphology of the dust-traced IVS and the Gum Nebula's $\rm H\alpha$ emission when projected onto the sky. We quantify contributions of feedback from stellar winds, an expanding HII region, and supernovae to the IVS formation, finding that stellar winds are subdominant. Our momentum analysis shows that both an HII region and supernova feedback could drive the shell's expansion. Using astrometric constraints from Gaia and Hipparcos, we trace back nearby feedback sources and find that the massive stars $\gamma2$ Velorum and $\zeta$ Puppis are currently within the IVS, producing enough ionizing luminosity to form an HII region of comparable size. Alternatively, if the IVS' momentum is primarily driven by supernovae, $1-2$ events would be required. We also identify several young massive clusters that could have hosted supernovae within the past 3 Myr.
\end{abstract}



\section{Introduction} \label{sec:intro}
\begin{figure*}
    \centering
    \includegraphics[width=1\textwidth]{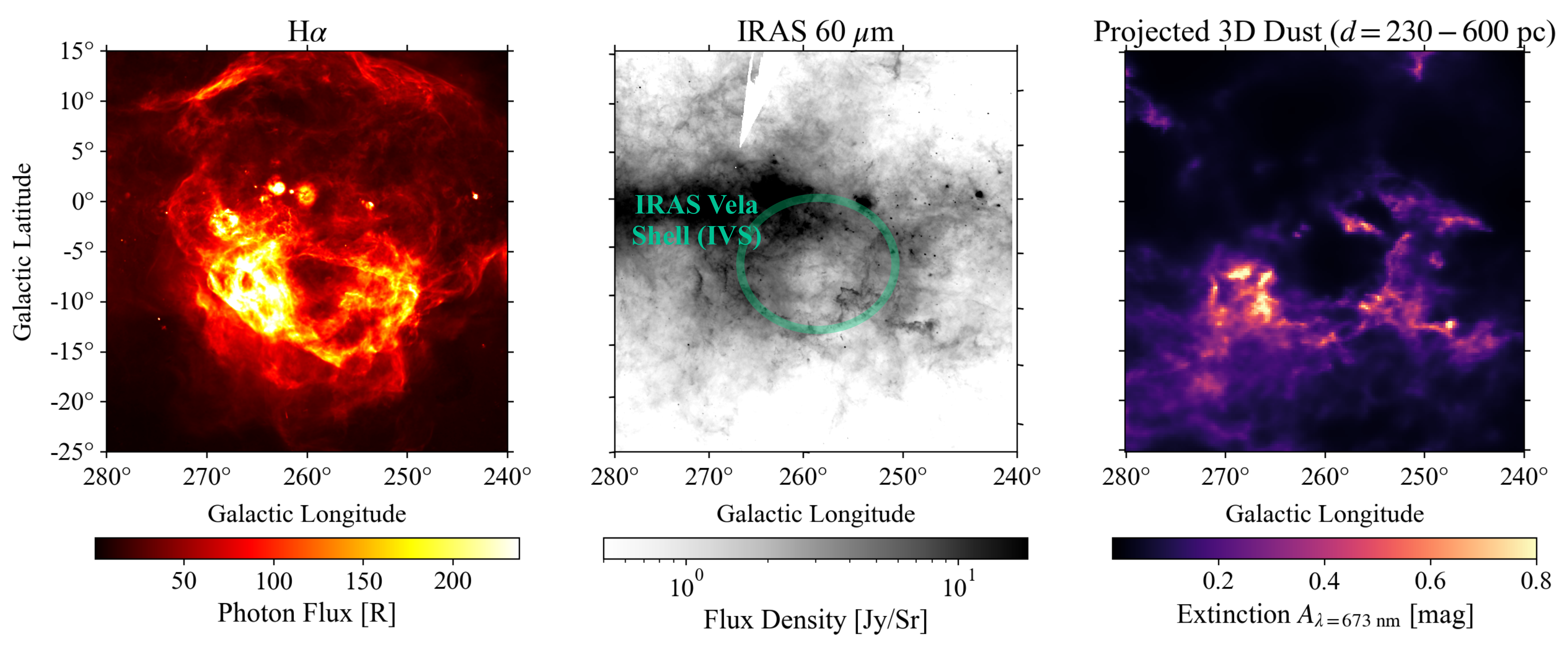}
    \caption{The Gum Nebula region as seen in different tracers. \textit{Left}: H$\alpha$ emission \citep{Finkbeiner2003}. \textit{Middle}: IRAS 60 $\mu$m emission \citep{Miville-Desch2005}, where the prominent ring-like structure is the IRAS Vela Shell (IVS). \textit{Right}: projected 3D dust map from \citet{Edenhofer2024} (converted to Gaia G-band extinction) and integrated over the range of distances consistent with the IVS  ($d = 230-600 \; \rm pc$).}
    \label{fig:gum_threepanel}
\end{figure*}

The Gum Nebula is one of the most prominent regions of H$\alpha$ emission in the southern sky, first recognized by \citet{Gum1952}. Located at a distance of roughly 450 pc \citep{Brandt1971, Eggen1980} from the Sun, the nebula spans over $30^\circ$ on the plane of the sky (see left panel of Figure \ref{fig:gum_threepanel}). The physical properties of the nebula have been widely studied, including the mean electron density ($1.3\;\rm cm^{-3}$), angular radius ($22.7^\circ$), thickness ($18.5\;\rm pc$) \citep{Purcell2015}, and temperature (11,300 K) \citep{Reynolds1976I}. Evidence of feedback processes pervades this region \citep[e.g.,][]{Sridharan1992, SridharanThesis1992, Bhatt1993, Woermann2001, 2006Testori, 2019Cantat-Gaudin}, as evidenced by the Vela supernova remnant \citep{Duncan1996}, the fast-moving runaway O star $\zeta$ Puppis \citep{Pauldrach1994, Schaerer&Schmutz1994, Howarth2019}, the binary stellar system $\gamma^2$ Velorum (containing a Wolf-Rayet star) \citep{Reynolds1976, DeMarco1999}, and a collection of young massive clusters, including the Vela OB2 association \citep{Kapteyn1914, deZeeuw1999} and Trumpler 10 \citep{deZeeuw1999, Hunt2023}. 

These sources of potential feedback appear to have collectively shaped the evolution and present-day structure of the surrounding interstellar medium. For example, the Gum Nebula encompasses a system of over thirty compact molecular globules known as cometary globules \citep[CGs;][]{Hawarden1976, Zealey1983, Reipurth1983}. These CGs are distributed over a ring-like shape, featuring a head-tail morphology with denser, bright-rimmed heads that sometimes contain star formation \citep{Pettersson1987,Sahu1992, SridharanThesis1992, Reipurth1993, Kim2005,Yep_CG30_2020}, and fainter tails that appear to point away from a common center. Consistent with their morphologies, the system of CGs also appear to be dynamically expanding away from a common center. This expansion was first observed by \citet{Sridharan1992, SridharanThesis1992} through $^{12}\rm{CO}$ radial velocity observations.

A structure known as the IRAS Vela Shell (IVS) also overlaps the Gum Nebula on the plane of the sky (see Figure \ref{fig:gum_threepanel}). The IVS spans $15^\circ$ in the sky and has been extensively studied with a variety of tracers. \citet{Sahu1992thesis} investigated the IVS with the infrared IRAS Sky Survey Atlas (ISSA) at $25$, $60$ and $100 \; \mu$m (see middle panel of Figure \ref{fig:gum_threepanel}), and gave the structure its name. The IVS has been traced in atomic gas, molecular gas, and in the previously identified, expanding system of molecular CGs that appear to trace its edge in projection, finding the IVS to be expanding at a velocity of roughly $8-13 \; \rm km \; s^{-1}$~\citep{Sahu1993, Rajagopal1998}. A range of explanations for the CGs system and the IVS expansion have been put forward, including stellar winds from an OB association Vela OB2 \citep{Rajagopal1998} and a potential supernova event that could have carved out the IVS and triggered the formation of the Vela OB2 cluster~\citep{2006Testori, 2019Cantat-Gaudin}.


A central focus of these studies has been the physical and dynamic relationship between the Gum Nebula, the IVS, and the system of CGs. The question of whether they are co-spatial or separate structures located at different distances has been long debated. For example, \citet{Sahu1993} performed a spectroscopic study of the ionized gas in the vicinity of the Gum Nebula, finding that although the IVS is expanding, there is no corresponding systematic motion in the ionized gas associated with the Gum Nebula, and thus the IVS and Gum Nebula are separate structures. 
\citet{Rajagopal1998} concurred that the IVS and Gum Nebula are unrelated, although they concluded that CGs are indeed part of the IVS. The notion that IVS and the Gum Nebula are unrelated was challenged by \citet{Woermann2001}, who investigated the neutral material around the Gum Nebula using hydroxyl emission lines. They found no significant differences between the mid-IR to radio flux ratio in these two regions, suggesting that IVS is not a separate structure.

Despite the range of interpretations on the nature of the IVS and its relationship to the Gum Nebula, past studies have been limited to either 2D plane-of-sky projections, or spectral-line mapping, where the third dimension is the radial velocity of the gas. With the advent of high-precision astrometry, photometry, and spectroscopy from the Gaia space telescope \citep{2023GAIAbprp}, alongside complementary photometric surveys \citep[e.g. 2MASS, WISE][]{Skrutskie_2006,Wright_2010}, it is now possible to resolve the three-dimensional (3D) spatial distribution of interstellar gas in the solar neighborhood thanks to a technique called 3D dust mapping \citep[for reviews, see e.g.][]{Zucker2023}. With this technique, it is possible to create 3D maps of the IVS at 1 pc distance resolution (see the right panel of Figure \ref{fig:gum_threepanel} for a view of the IVS as seen over a narrower range of distances accessible with 3D dust mapping). Simultaneously, Gaia provides new constraints on the 3D spatial positions and motions of feedback sources, including, for example, the system of young massive clusters overlapping the IVS on the plane of the sky. The combination of the two offers an unprecedented opportunity to connect the 3D spatial structure of the IVS and the broader Gum Nebula region with the sources of feedback potentially powering its expansion.

In this study, we aim to leverage this new high-resolution 3D dust map to constrain the 3D geometry of the IVS, connecting this new geometry with new constraints on the 3D distribution and motion of sources of feedback to understand the shell's formation and evolutionary history. We will also provide new evidence that the Gum Nebula and the IVS are physically associated with each other in 3D space. 

First, we describe the 3D dust map data utilized in this work (\S\ref{sec: data}) alongside our methodology for constraining the 3D geometry of the IVS (\S\ref{sec:Methods}). In \S\ref{sec: result} we describe the physical properties of the IVS derived from our new 3D geometric model, including its mass, density, and size. Combining our geometric model with extant constraints on its expansion from the literature, in \S\ref{sec: discussion} we discuss the dynamical properties of the IVS, including estimates for its momentum and energy. Using these estimates, we characterize potential contributions from stellar winds, HII region, and supernovae to the shell's formation, in the context of the 3D distribution and dynamics of nearby massive stars and clusters. Finally, we present our conclusions in \S\ref{sec: conclusion}.

\section{Data} \label{sec: data}
High-precision astrometry, photometry, and low-resolution spectroscopy have inspired the creation of a number of 3D dust maps \citep[see e.g.][]{2019Green, Leike2020, 2022Lallement, 2022Leike, Edenhofer2024}. Leveraging the reddening effect that dust has on the colors of stars (whose distances are constrained with Gaia), these maps chart the local dust distribution in three spatial dimensions, resolving the ISM structure over a range of densities. We utilize the \citet[hereafter, E24]{Edenhofer2024} 3D dust map for our study of the IVS, as it offers both high spatial resolution and complete coverage of the Gum Nebula region. The E24 map reconstructs the 3D distribution of dust out to a distance of 1.25 kpc from the Sun with an angular resolution of 14$'$ and a distance-dependent spatial resolution varying between 0.4 pc at close-by distances and a few pc at far away distances. Since the Gum Nebula is located $\sim450$ pc away from the Sun in the southern sky, this map is ideal for our study. 

The E24 dust map uses a novel, scalable statistical model for spatial smoothness \citep{Edenhofer2022} and a newly developed computational framework to bring the inference derivation onto the GPU \citep{Edenhofer2024_software}. The map leverages the distance and extinction estimates of 54 million stars provided by the catalog of \citet{Zhang2023}. The latter work uses the Gaia DR3 BP/RP spectra~\citep{2023GAIAbprp}, supplemented by 2MASS and WISE photometry \citep{2MASS2006, 2010WISE}, to derive a forward model that estimates stellar atmospheric parameters, distances, and extinctions. \citet{Edenhofer2024} released 12 samples of the dust map, along with the mean and standard deviation of these samples. The map provides estimates of the dimensionless extinction density (hereafter $A_{\rm ZGR}'$), as a function of 3-D location. The variable $A_{\rm ZGR}'$ can be converted to differential extinction at any arbitrary wavelength given the \citet{Zhang2023} published extinction curve. To convert from dimensionless extinction density to the total volume density of hydrogen nuclei, we adopt the formalism from \citet{O'Neill2024} (see their Appendix D for full derivation): 

\begin{equation} \label{eqn:H}
n_{\rm H} =1653 \cdot A_{\rm ZGR}' \; \rm{cm}^{-3},
\end{equation}
which assumes a constant extinction to column density ratio based on \citet{2011Draine}. This volume density represents the total density of hydrogen nuclei, regardless of phase ($n_{\rm H} = n_{\rm H^{+}} + n_{\rm HI} + 2\times n_{\rm H_2}$) and is used to derive all the results reported in \S\ref{sec: result} \citep[see also conversion from differential extinction to hydrogen volume density from][]{Bialy2021}.

\section{Methods} \label{sec:Methods}
In this section, we describe our methodology for deriving the 3D geometry of the IVS from the E24 3D dust map. This includes the methodology for identifying the inner and outer shell boundaries, shell thickness, and the peak $n_{\rm H}$ density and its corresponding radius (\S \ref{subsec: boundary}). Using the 3D geometry information, we also derive the mass enclosed within the shell (\S \ref{subsec: mass}). We address the uncertainty in our calculations in \S\ref{subsec: uncertainty}.

\subsection{Deriving the 3D Shell Geometry} \label{subsec: boundary}
To derive the 3D geometry of the IRAS Vela Shell (IVS), we first create a series of {\tt HEALPix} \citep{2005Healpix} spheres with a grid resolution of $N_{\text{side}}=128$, spaced at 1 pc intervals in radius, and all centered at $(x,y,z)=(-67, -334, -65)$ pc in Heliocentric Galactic Cartesian Coordinates. This setup provides an angular resolution of $27'$ and a spatial resolution of at least 2.8 pc for a sphere at 350 pc, with finer resolutions closer to the IVS center. The center of the shell was identified by a visual inspection of the E24 dust map, rendered in 3D using the {\tt glue} visualization software \citep{2019glueviz}. Based on the literature, we searched for a shell-like structure in the region spanned by $l\in[250, 270]^\circ, b\in[-20,0]^\circ$ \citep{Sahu1992, Woermann2001} and at a distance of $d\in[300,450]$~pc from the Sun \citep{Sahu1992, 2011Sushch}. We verified that our results reported in \S\ref{sec: result} remain consistent when adopting a reasonable range of values for the current central point within 15 pc, as long as the center is visually located within the shell interior. 

We then generate a series of rays connecting the central point to every pixel on the outer HEALPix sphere at $r=350 \rm \; pc$, which fully encompasses the 3D extent of the potential shell-like structure in the E24 map. Along each ray, we query the extinction density at 1 pc intervals, corresponding to the same HEALPix point for each layer of the spheres. Applying the conversion factor from Eq.~\ref{eqn:H}, we obtain $n_{\rm H}$ along each ray. Thus, each ray captures the interpolated density between the center of the sphere and the maximum radial distance we consider (350 pc). To minimize the effects of noise, we smooth the density profile using a Gaussian smoothing kernel $\sigma_{\rm smooth}$. The choice of the kernel primarily affects the resulting shell's thickness (broadening or narrowing the inner and outer shell boundaries) with minimal effects on the location of the peak density and radial distance, as defined below. Further discussion on our choice of Gaussian smoothing kernel is provided in Appendix \ref{appendix: kernel} and incorporated into our uncertainty estimates in \S \ref{subsec: uncertainty}. For the remainder of this work, we adopt $\sigma_{\rm smooth}=10$~pc as our fiducial smoothing value.

\begin{figure*}[]
    \centering
        \includegraphics[trim={0 0 0 0}, width=1.05\columnwidth]{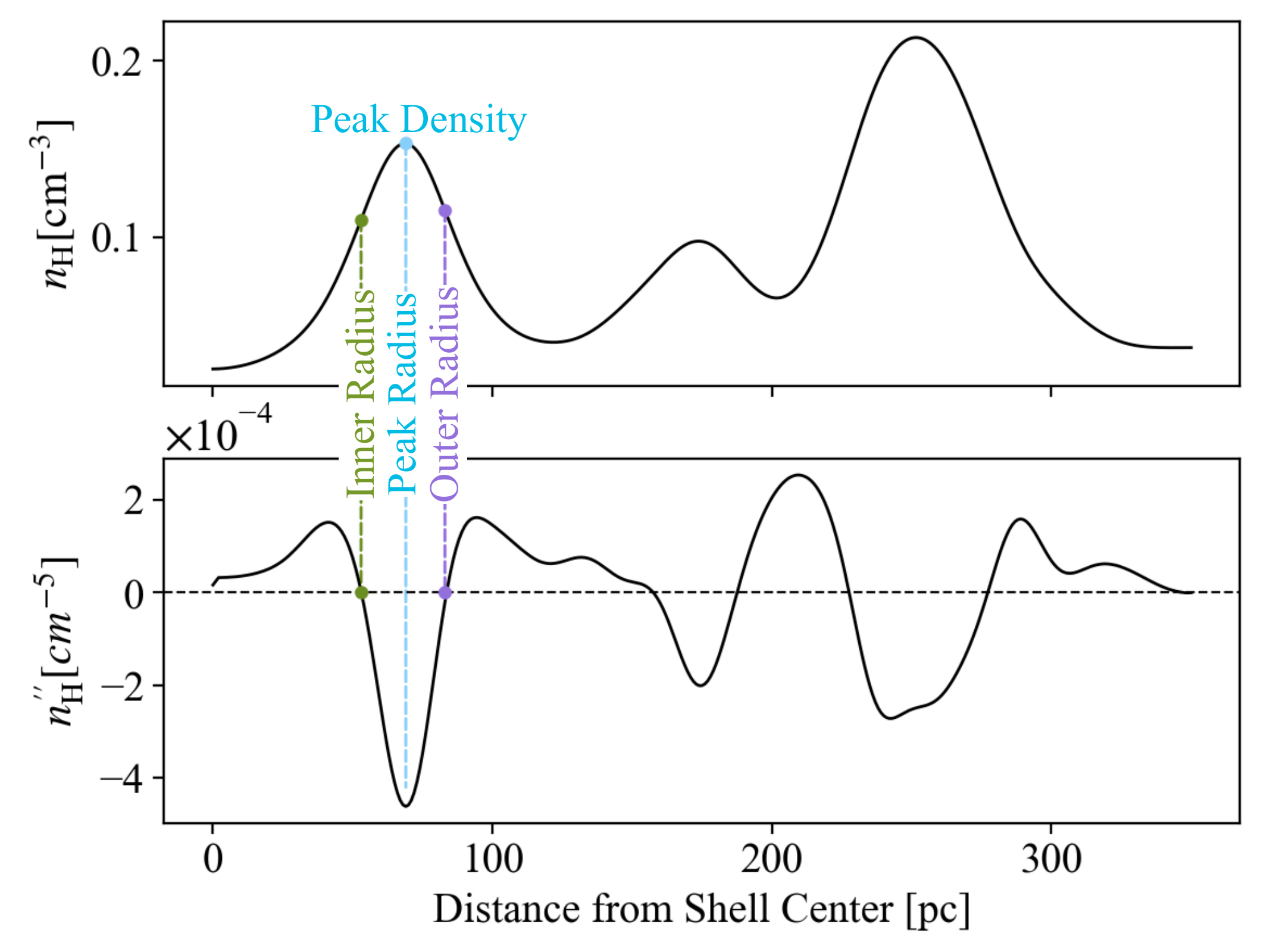}
        \includegraphics[trim={0 1cm 0 0},width=1.05\columnwidth]{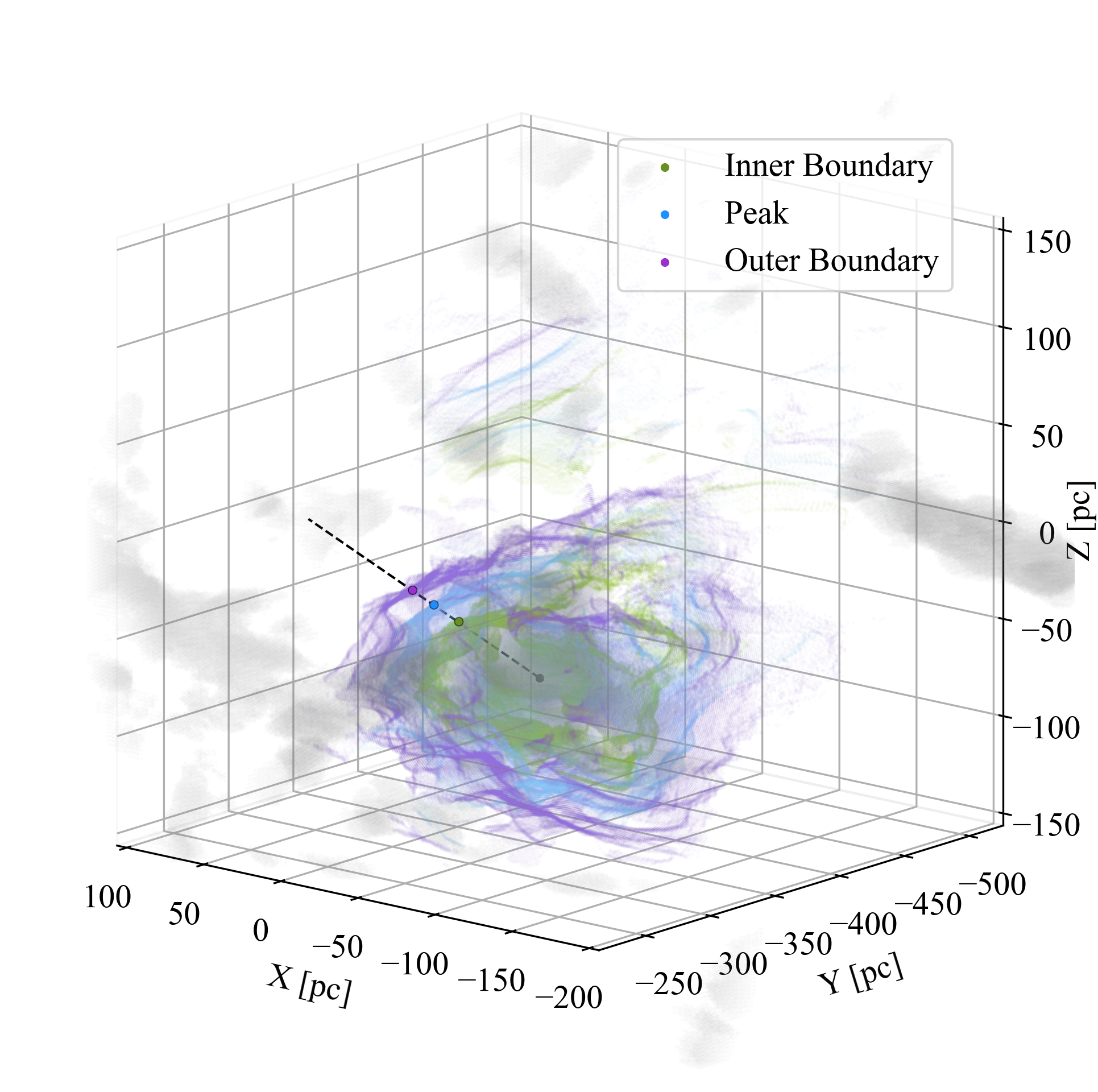}
    \caption{Derivation of the IVS shell geometry. \textit{Left}: the upper panel displays an example of density profile $n_{\rm H}$, while the lower panel shows the second derivative $n''_{\rm H}$ which is used to derive the shell boundaries. \textit{Right}: inner shell radius (green), peak shell radius (blue), and outer shell radius (purple) of the IVS as derived in this work. The dashed line represents an example ray, which corresponds to the density profile shown on the left. For the rays exhibiting multiple peaks, our method (see Eqns.\ref{eqn: innner bound}, \ref{eqn: outer bound}) selects the first peak as the shell boundary. }
    \label{fig:density}
\end{figure*}

Our method for finding the inner and outer IVS boundaries is adapted from \citet{2020Pelgrims}, who derived the 3D geometry of the Local Bubble (the supernovae-driven cavity around the Sun) using inflection points in the extinction density distribution. In this work, we compute the first ($n_{\rm H}'$) and second ($n_{\rm H}''$) derivatives of the hydrogen volume density with respect to the distance from the center, $r$, to locate the shell boundaries along each ray.

We define the radial distance of the IVS inner boundary $r_{\rm in}$ to be the first inflection point where the density profile changes from convex to concave:
\begin{equation}\label{eqn: innner bound}
    r_{\rm{in}} = {\rm arg\; min}_{r} \{r | n''_{\rm{H}}(r)=0, n'_{\rm{H}}(r)>0\},
\end{equation}
where $n'_{\rm{H}}$ and $n''_{\rm{H}}$ are the first and second derivatives of $n_{\rm H}$ with respect to distance $r$ from the center, respectively. 
For the outer boundary, due to the mild irregular fluctuations where inflections points may not clearly signal density changes, we select the first point where the second derivative changes sign (from negative to positive or vice versa) at a radius larger than inner boundary. This criterion can be written as: 
\begin{equation}\label{eqn: outer bound}
    r_{\rm {out}} = {\rm arg\; min}_{r} \{r | r>r_{\rm in}, n''_{\rm{H}}(r)=0\}.
\end{equation}

In most cases, once the inner boundary is determined, the outer boundary corresponds to the point where the second derivative is zero and the first derivative is negative: $ r_{\rm {out}} = {\rm arg\; min}_{r} \{r | r>r_{\rm in}, n''_{\rm{H}}(r)=0, n'_{\rm{H}}(r)<0\}$. For cases where the second derivative does not successfully capture the second inflection point of the density profile, our method still reliably identifies an outer boundary that reflects density variations. Across all samples of the E24 map to which we applied this algorithm, the inner boundary was consistently found along each ray. The number of rays with an unrecoverable outer boundary was generally fewer than 10 (out of 196,608) for each sample of the E24 map.


An example of the density profile as a function of radial distance from the IVS center is shown in the left panel of Figure \ref{fig:density}. The upper panel displays the $n_{\rm H}$ profile as a function of distance to the IVS center, while the lower panel shows $n''_{\rm H}$. The inner and outer boundaries are demarcated in both panels. We define the shell thickness $\Delta_{\rm shell}$ to be the distance between $r_{\rm in}$ and $r_{\rm out}$. 

We define the peak of the shell to be the first local density maximum between the inner and outer boundaries, which is less sensitive to $\sigma_{\rm smooth}$ than the boundaries (see Appendix \ref{appendix: kernel} for further discussion). The peak density value $n_{\rm peak}$ and its corresponding radial distance $r_{\rm peak}$ for the example ray are also labeled in the left panel of Figure~\ref{fig:density}. In the right panel, we show the resulting boundary and peak locations for the entire IVS.

We verified our results using the newly published method from \citet{O'Neill2024}, which leverages a peak-finding method to constrain the Local Bubble's geometry. The application of this method to the IVS and a comparison of the two approaches is discussed in Appendix \ref{appendix: method-compare}. The primary difference between the two methods lies in the quantification of the left and right endpoints of the curve. The 3D geometry of the IVS remains qualitatively consistent between the two methods.

\subsection{Deriving the Shell Mass}\label{subsec: mass}
To calculate the IVS mass along each ray, we integrate $n_{\text{H}}$ between $r_{\rm{in}}$ and $r_{\rm{out}}$ with a step size of 1 pc along each density profile:
\begin{equation}\label{eqn: Mray}
    M_{\rm{ray}} = \sum_{i} {\rm d}M_{i} = \sum_{i,\rm{in}}^{i,\rm{out}} 1.37m_{\rm p}\times n_{{\rm H},i}\times {\rm d}V_{i} , 
\end{equation}
where $n_{\text{H},i}$ is the hydrogen number density in the $i$th bin along one ray, and $1.37m_{\rm p}$ accounts for the mean particle mass, including the contribution of helium assuming solar abundance \citep[c.f.][]{Zucker2021,Bialy2021}. Given the 1 pc sampling of our density profile, we further expand ${\rm d}V_{i}$ as follows:
\begin{equation}
    {\rm d}V_{i} =  \frac{4\pi r_{i}^2}{N_{\text{pix}}}dr \\
           = \frac{1}{N_{\text{pix}}}\times \frac{4\pi}{3} (r_{i+1}^3 -r_i^3)
           \label{eq:volume_element}
\end{equation}
where $N_{\text{pix}} = 196,608$ is the total number of pixels on the HEALPix surface at $N_{\text{side}}=128$.

\subsection{Uncertainties} \label{subsec: uncertainty}
In our analysis of the shell's geometry and its derived properties, we account for both statistical uncertainties arising from the 3D dust map and systematic uncertainties arising from our choice of smoothing kernel $\sigma_{\rm smooth}$.

The statistical uncertainty primarily stems from the uncertainties inherent in the underlying 3D dust map, captured by the twelve samples released in \citet{Edenhofer2024}. To quantify this, we adopt a fiducial smoothing kernel $\sigma_{\rm smooth}= 10$ pc and apply this fixed smoothing kernel to all twelve dust map samples, deriving twelve IVS models that capture the statistical uncertainty. 

Systematic uncertainty, on the other hand, stems from the choice of smoothing kernel $\sigma_{\rm smooth}$, since larger smoothing kernels broaden the volume density distribution along each ray, while smaller smoothing kernels make our approach more sensitive to noise. To quantify this effect, we fix the 3D dust map to the released posterior mean map from E24 and vary $\sigma_{\rm smooth}$ in 1 pc intervals between 6 to 14 pc, deriving nine IVS models capturing the systematic uncertainty. In Appendix \ref{appendix: kernel}, we provide examples that illustrate how the choice of smoothing kernels affects the resulting boundaries. 

When calculating uncertainties on all properties reported in \S \ref{sec: result}, we compute the statistical and systematic uncertainties given these sets of models, and add the statistical and systematic uncertainties in quadrature to derive the total uncertainty as standard deviation: $\sigma_{\rm{total}} \equiv \sqrt{\sigma_{\rm{sys}}^2 + \sigma_{\rm{stat}}^2}$. To enable follow-up investigations, we release both the fiducial model and the models derived with different smoothing kernels $\sigma_{\rm smooth}$ and dust map realizations at the Harvard Dataverse (see \url{ https://doi.org/10.7910/DVN/VZMYSL} and \url{https://doi.org/10.7910/DVN/TVCC1H} respectively). 

\section{Results} \label{sec: result}
In this section, we present the 3D geometry of the IVS along with its physical properties. In Table~\ref{tab:fiducial model}, we summarize the variation in shell properties across the shell's surface for our fidicual model. In Table ~\ref{tab:property}, we summarize the derived properties of the shell (mass, momentum, etc.).

\begin{deluxetable}{cccccc}
\tablecaption{Summary of Geometric Properties of the IVS from the Fiducial Model \label{tab:fiducial model} }
\colnumbers
\tablehead{\colhead{$r_{\rm in}$} & \colhead{$r_{\rm out}$} & \colhead{$\Delta_{\rm shell}$} & \colhead{$r_{\rm peak}$} & \colhead{$n_{\rm H,\; peak}$} & \colhead{$d_{\rm IVS}$} \\
\colhead{pc} & \colhead{pc} & \colhead{pc} & \colhead{pc} & \colhead{$\rm cm^{-3}$} & \colhead{pc} }
\startdata
$49_{-20}^{+33}$ & $82_{-35}^{+34}$ &  $31_{-22}^{+28}$ & $70_{-22}^{+84}$ & $0.31_{-0.28}^{+11.35}$ & $353_{-77}^{+77}$
\enddata
\tablecomments{(1) Inner boundary radii (2) Outer boundary radii (3) IVS thickness (4) Peak radius (5) Peak density (6) Distances between the pixels on the IVS surface and the Sun. Each quantity's density distribution is shown in Figure \ref{fig:shell_property}. The listed values represent the median values (indicated by black dashed vertical lines in Figure \ref{fig:shell_property}) and 95\% percentile ranges (indicated by green dashed vertical lines in Figure \ref{fig:shell_property}). As these properties are derived from the fiducial model, the lower and upper bounds capture the variation in each property across the shell's surface.}
\end{deluxetable}

\subsection{3D Geometry of the Shell}
\begin{figure*}
    \includegraphics[width=1.\textwidth]{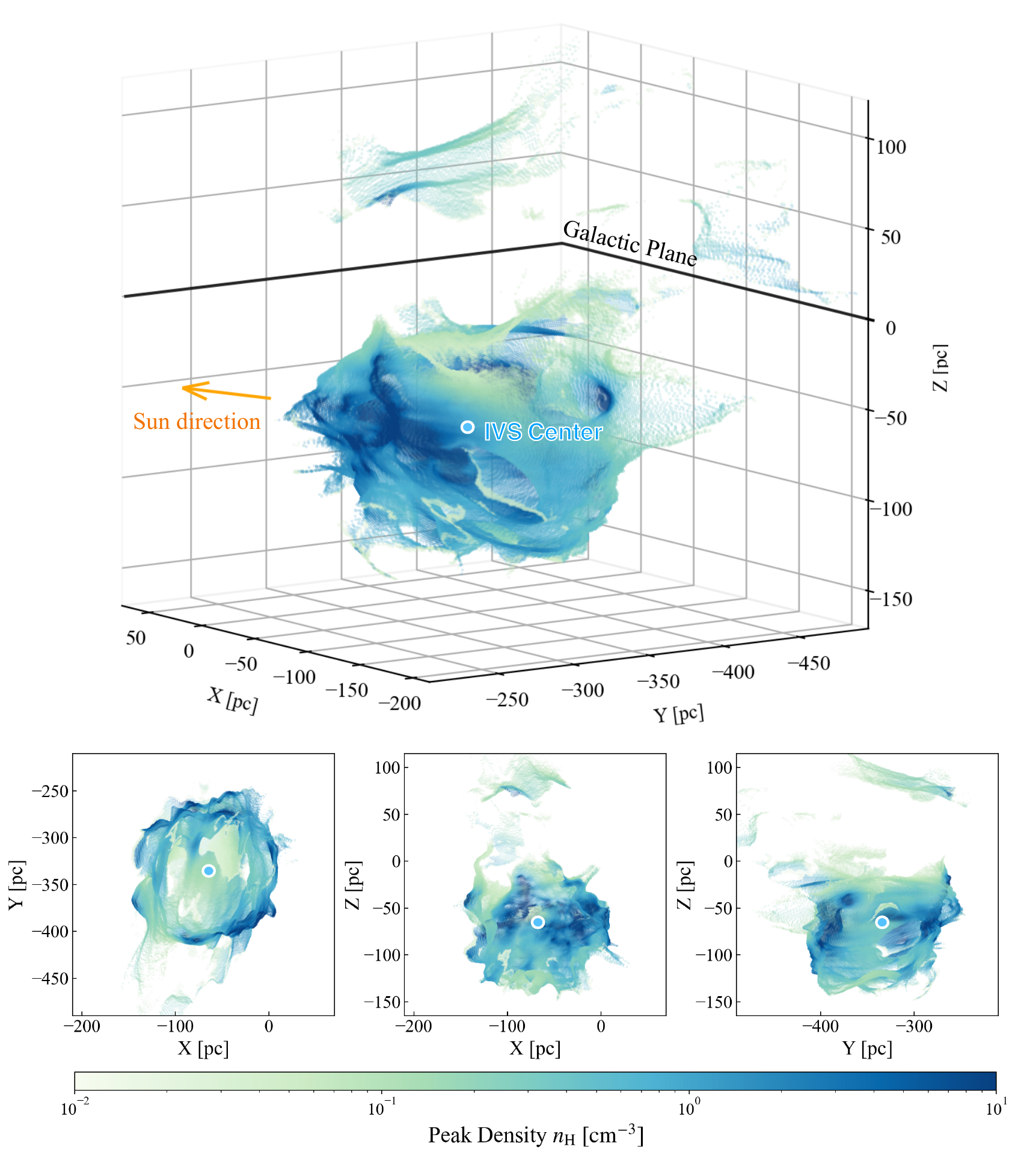}
    \caption{3D geometry of the IVS, defined by the radial distance from the center to the density peak along each ray. The surface is color-coded by peak density value $n_{\rm H}$, spanning four orders of magnitude in dynamic range. The orange vector points towards the Sun.}
    \label{fig:peak3D}
\end{figure*}

\begin{figure*}
    \includegraphics[width=1.05\textwidth]{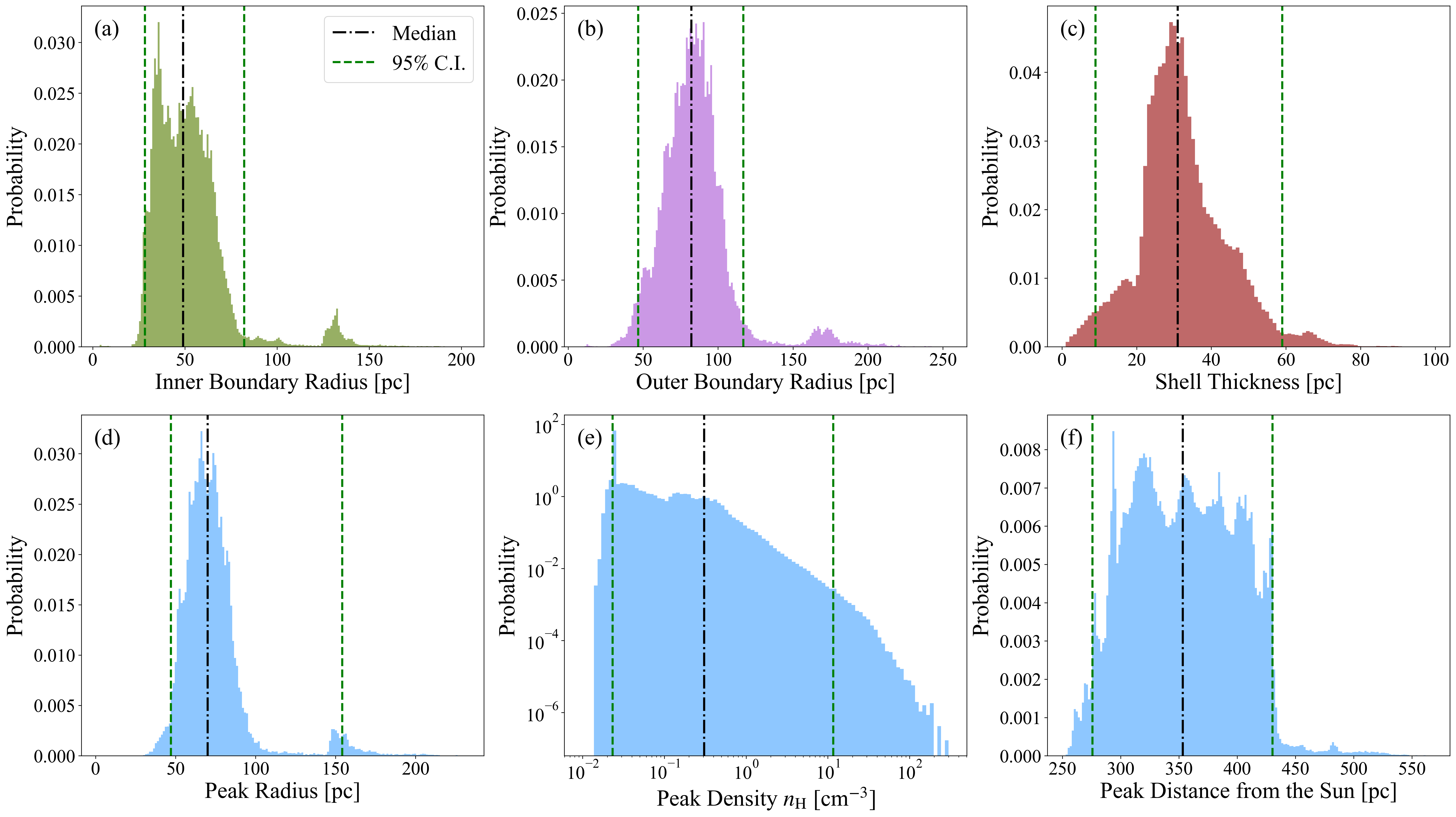} 
    \caption{Probability density distributions of the IVS properties based on our fiducial model. \textit{Top panels from left to right}: the inner boundary radii distribution from the geometric center; the outer boundary radii distribution from the geometric center; shell thickness. \textit{Bottom panels from left to right}: peak radii distribution from the geometric center; the distribution of $n_{\rm H}$ values at each peak location; distance from the Sun of the peak radii distribution.}
    \label{fig:shell_property}
\end{figure*}

In Figure \ref{fig:peak3D} we show our fiducial model for the 3D geometry of the IVS (defined by the peak radius) colored according to the peak density. The model features a relatively dense ``bowl"-like structure around 70 pc below the Galactic plane and a more diffuse filament approximately 70 pc above it. As observed in Figure \ref{fig:peak3D}, the section of the shell closer to the Sun is denser. This enhanced density may be due to an interaction between the IVS and the Local Bubble \citep[c.f.][]{O'Neill2024}, which shares a wall with the IVS on the IVS's near side. The PDFs of the shell properties for our fiducial model are shown in Figure \ref{fig:shell_property} (see also Table~\ref{tab:fiducial model}). We find a typical peak radius of $70$ pc with a spread of $_{-22}^{+84}$ pc over the surface. The corresponding peak density spans roughly four orders of magnitude, ranging from $10^{-2} \;\rm cm^{-3}$ to $10^2\; \rm cm^{-3}$. The typical shell thickness is $31$ pc (spread of $_{-22}^{+28}$ pc). The IVS (as defined by the distribution of peak radii) is located at a typical distance of $\sim353$ pc from the Sun as shown in the lower right panel of Figure \ref{fig:shell_property}, considerably closer than previous studies placing it at $\sim$450 pc \citep{1993Sahu_ivs_structure}. The properties in Table \ref{tab:fiducial model} are calculated from our fiducial model, the uncertainties are \textit{not} akin to the uncertainties discussed in \S \ref{subsec: uncertainty}, but rather capture the \textit{variation} in that property across the surface of the shell.

\begin{deluxetable}{ccccc}
\tablecaption{Derived Physical Properties of the IVS (95\% C.I.) \label{tab:property} }
\colnumbers
\tablehead{\colhead{Mass} & \colhead{Momentum} & \colhead{Energy} & \colhead{$n_{\rm {H, ambient}}$} & \colhead{$n_{\rm{H, interior}}$ } \\
\colhead{$\times 10^4\rm M_{\odot}$} & \colhead{$ \times 10^5\rm M_{\odot}\; km\; s^{-1}$} & \colhead{$ \times 10^{49}$erg} & \colhead{$\rm cm^{-3}$} & \colhead{$\rm cm^{-3}$} }
\startdata
$5.1_{-2.4}^{+2.4} $ & $6.0_{-3.4}^{+4.7}$ & $7.1_{-5.4}^{+10.8} $ & $0.67_{-0.26}^{+0.26}$ & $0.17_{-0.12}^{+0.12}$
\enddata
\tablecomments{(1) Total mass of the IVS (2) Total momentum of the IVS (3) Total shell kinetic energy (4) Predicted ambient density prior to shell formation (5) Mean present-day shell interior density. The values are derived based on our fiducial model, with the uncertainties obtained following \S\ref{subsec: uncertainty}.}
\end{deluxetable}

\subsection{Shell Momentum and Kinetic Energy}\label{subsec:momentum_energy}
We combine our total shell mass, $M_{\rm IVS}=(5.1 \pm 2.4) \times 10^{4}\; \rm M_{\odot}$, derived using the formula in \S\ref{subsec: mass} (uncertainties calculated according to \S\ref{subsec: uncertainty}), along with an adopted shell expansion velocity, to determine the kinematic properties of the IVS, specifically its expansion momentum and energy. We assume that the average shell velocity follows a Gaussian distribution (with $1\sigma$ uncertainty) of $v_{\rm exp} = 12\pm 3\;\rm km\;s^{-1}$ from \citet{Sridharan1992} and \citet{Rajagopal1998} who derived this value by modeling the $^{12}\rm CO$ radial velocities towards the system of cometary globules. This velocity range is consistent with other measurements in the literature, including the NII emission line studied by \citet{Sahu1993} and $^{12}\rm CO$ observations in the entire IVS region by \citet{Rajagopal1998}. \citet{Woermann2001} obtain different expansion rates for the front and back sides of the shell, at $14 \rm \; km \; s^{-1}$ and $8.5\rm \; km \; s^{-1}$ respectively, which is consistent with the velocity range we consider here. We note that this velocity range may not perfectly match the top filament due to density differences between the two sections. However, in the absence of separate velocity measurements, we apply the same $v_{\rm exp}$ to the entire IVS structure. As we will demonstrate in \S\ref{subsec: IVS_Gum_CG}, a subset of the CGs studied in \citet{Sridharan1992} lie at a distance and possess a radial velocity consistent with both our fiducial shell model and the radial velocity of the stars embedded in CG30. Therefore, we find multiple evidences supporting the adoption of this velocity range. 

Assuming spherically symmetric radial expansion, the total shell momentum, $p$, is given as:

\begin{equation}
    p = M_{\rm IVS}\cdot v_{\rm exp},
\end{equation}
while the shell kinetic energy, $E_k$, is given as:  
\begin{equation}
    E_{k} = \frac{1}{2} \cdot M_{\rm IVS}\cdot v_{\rm exp}^{2}.
\end{equation}
We draw 10,000 samples from the Gaussian distribution of $M\rm_{IVS}$ and $v_{\rm exp}$ to estimate the uncertainties in the shell's momentum and energy. We quote the resulting median and 95\% C.I. values in Table~\ref{tab:property}. Our energy estimate of $7.1_{-5.4}^{+10.8} \times 10^{49}\;\rm erg$ is consistent with the value calculated by \citet{2006Testori}, who estimated the IVS expansion energy to be $1.7\times 10^{50} \;\rm erg$.

\subsection{Predicted Ambient Density}\label{subsec: density_ambient}
To place a constraint on the ambient interstellar environment prior to the shell's formation, we divide the total shell mass by the volume enclosed by the outer shell boundary ($V_{\rm outer}$), such that
\begin{equation}
    n_{\rm H, ambient} = (M_{\rm IVS} + M_{\rm interior}) / V_{\rm outer},
\end{equation} 
where $M_{\rm interior}$ is the mass enclosed inside the inner boundary of the IVS. 

Following the same method as in \S\ref{subsec: uncertainty} to compute uncertainties, we find a predicted ambient density of $n_{\rm H, ambient}=0.67\pm 0.26 \;\rm{cm^{-3}}$ (95\% C.I.). This density is consistent with the typical density of the warm neutral medium \citep[][WNM]{2011Draine}, also discussed in \citet{Bialy2019}. However, it is more than twice the observed \textit{median} peak density in Figure \ref{fig:shell_property}, likely due to the clumpy nature of the ISM, where the densest portions of the shell are primarily concentrated within a dense, ring-like structure in bottom ``bowl'' section. We also note that the \textit{mean} peak density of the skewed distribution is $1.38\;\rm{cm^{-3}}$. We can compare this predicted density to the present-day average interior density, which is given by
\begin{equation}
    n_{\rm H,interior} = M_{\rm interior} / V_{\rm inner}.
\end{equation}
We find $n_{\rm H,interior}=0.17\pm0.12\;\rm{cm^{-3}}$ ($95\%$ C.I.), indicating that the interior of the IVS retains a non-negligible density. This value also suggests that the ambient density prior to the IVS expansion was approximately five times higher than the present-day average density of the bubble's interior.

\section{Discussion} \label{sec: discussion}
\begin{figure}[htb!]
    \includegraphics[width=.5\textwidth]{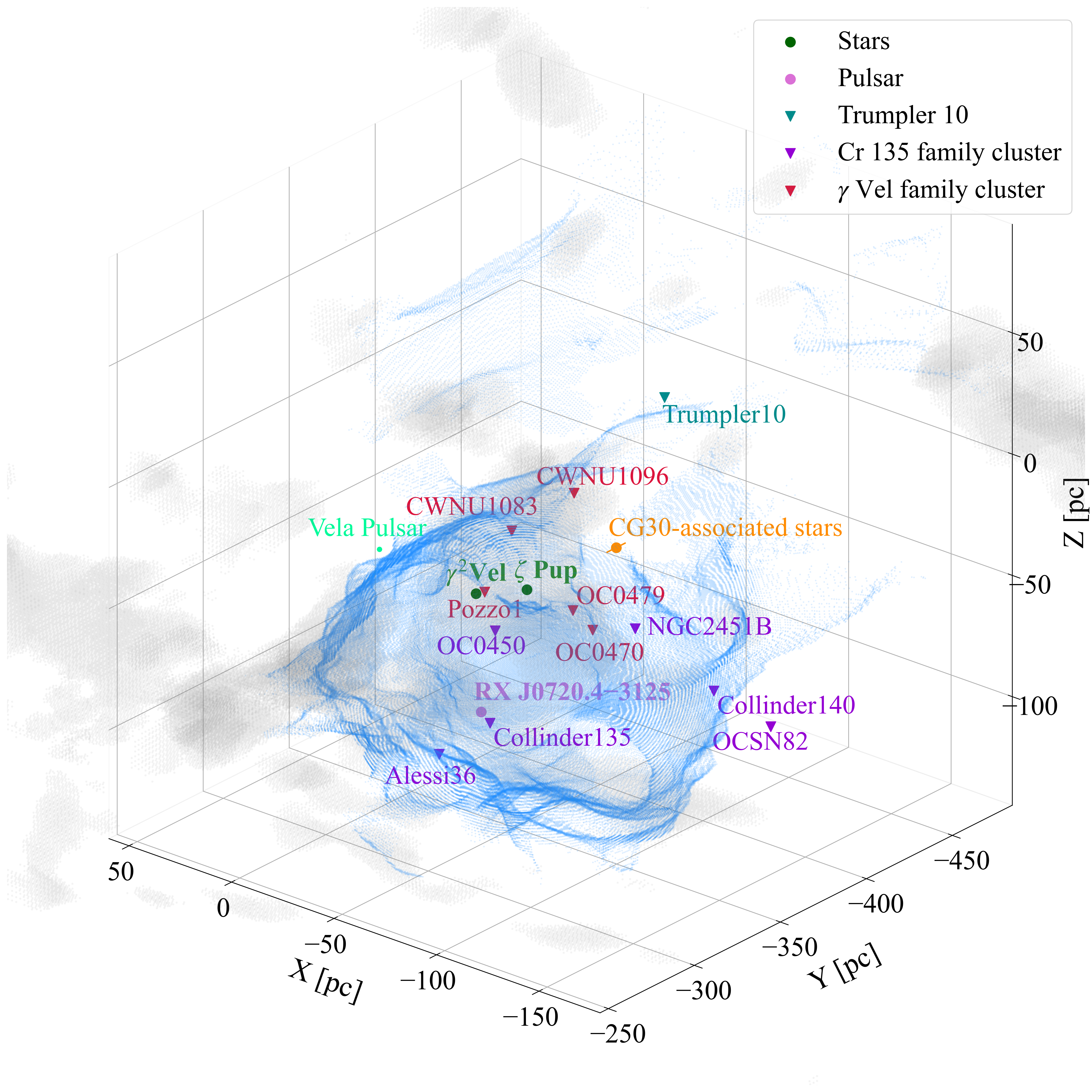}
    \caption{The fiducial model of the IVS (based on the peak radius) alongside potential nearby sources of recent feedback: $\zeta$ Puppis, $\gamma^2$ Velorum, the neutron star RX J0720.4-3125, and two sets of star clusters that belong to either the $\gamma$ Vel family or Cr 135 family \citep[c.f.][]{Swiggum2024}. A group of young stars associated with the cometary globule CG30 is marked in orange. The location of Vela pulsar \citep{Dodson2003} is marked in light green. An interactive version of this figure using \texttt{Plotly} \citep{plotly} is available \href{https://annie-bore-gao.github.io/Images/IVS_interactive.html}{here.} The interactive plot allows users to zoom in, rotate the 3D view, and examine the detailed relative positions of each component. Additionally, the side panel provides checkboxes to toggle the visibility of individual objects, enabling a clearer visualization of specific elements in the figure.}
    \label{fig:shell_w_stars}
\end{figure}

In this section, we contextualize the physical and dynamic properties of the IVS within its broader environment. We discuss the physical relationship between the IVS, the Gum Nebula, and the system of CGs. We also discuss the origin of the IVS by analyzing possible energy sources that may have contributed to its formation and expansion. In Figure \ref{fig:shell_w_stars}, we present an overview of key landmarks towards the region (including potential sources of feedback) that will be relevant to our discussion.

\subsection{The Spatial Relationship between the IVS, Gum Nebula, and the Cometary Globules} \label{subsec: IVS_Gum_CG}
As discussed in \S\ref{sec:intro}, the spatial relationship between the Gum Nebula, the IVS, and the system of CGs remains a topic of active debate. With the new IVS 3D geometry and properties obtained in \S\ref{sec: result}, we are in a unique position to revisit this question. We find evidence that the IVS, the Gum Nebula, and at least a subset of the CGs are spatially correlated. 

\begin{figure}
    \includegraphics[width=.5\textwidth]{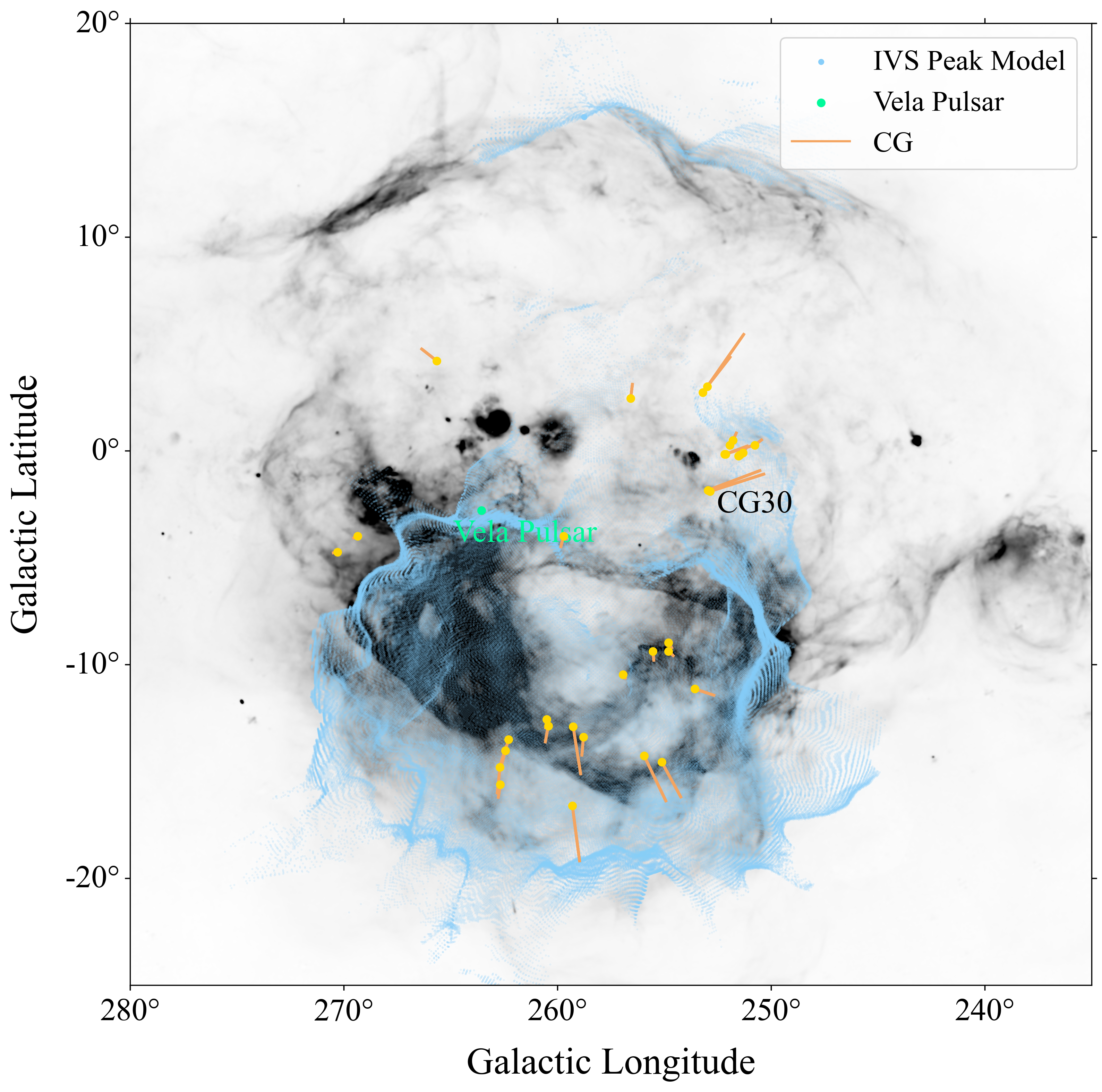}
    \caption{Spatial correlation between H$\alpha$ \citep[background grayscale][]{Finkbeiner2003}, the fiducial model of the IVS peak distribution, and the cometary globules \citep{Sridharan1992}. The CGs are represented by yellow heads and orange tails, with CG30 labeled.}
    \label{fig: Halpha_w_IVS}
\end{figure}

The Gum Nebula is typically traced by H$\alpha$ emission on the plane of the sky \citep[e.g.][]{Gum1952}. In Figure \ref{fig: Halpha_w_IVS}, we project our fiducial 3D IVS model onto the H$\alpha$ image of the Gum Nebula \citep{Finkbeiner2003}. We observe a strong correlation between the IVS peak distribution (in blue) and the projected morphology of the Gum Nebula, especially in the top filament. The bottom ``bowl'' of the IVS also aligns with the cavity seen in the H$\alpha$ intensity map. The spatial correlation between the dust emission and H$\alpha$ emission has also been observed towards the Orion-Eridanus superbubble \citep{Heiles1999}, and may reflect an ``onion-like" structure to the shell, with the neutral medium (traced by the dust) wrapping around the warm ionized medium (WIM; traced by the H$\alpha$).
Based on the strong morphological correlation between the H$\alpha$ and the IVS, the center of the Gum Nebula likely lies at distances $d < 450$ pc typically adopted in the literature, as the surface of the IVS lies at a median distance of $353 \pm 77$ pc from the Sun (see Figure~\ref{fig:shell_property} lower right panel).

We now turn to the relationship between the IVS and the CGs. The CGs form a ring-like structure (as seen projected on the sky in Figure \ref{fig: Halpha_w_IVS}) centered at $l=260.2^{\circ}, b=-4.1^{\circ}$ with a radius of $70-80$ pc \citep{Sridharan1992,Zealey1983, Reipurth1983} and have been regarded as tracers of the edge of the IVS \citep{Sridharan1992}. With the precise geometry of the IVS derived here, we can now connect the IVS and a CG directly by examining six stars embedded in CG30 as identified by \citet{Yep2020}. In their spectroscopic study, \citet{Yep2020} confirmed that six stars (PH$\alpha$ 14, PH$\alpha$ 15, KWW 464, KWW 598, KWW 1863, and KWW 2205) were born in the cometary globule CG30 and can serve as tracers of the dust in the CG30 region. The error-weighted mean LSR radial velocity for these six stars from \citet{Yep2020} is $v_{\rm lsr} = 7.1\pm 1.1\; \rm km \; s^{-1}$ (1$\sigma$ uncertainty), which is consistent with the radial velocity measured for the dense gas in CG30 based on CO emission, $v_{\rm lsr, CG30} = 5.8 \; \rm km \; s^{-1}$ by \citet{Sridharan1992}. 

We retrieve updated parallaxes for these stars from Gaia DR3 \citep{Gaia2023} and compare their mean distance to the 3D geometry of IVS. The stars are located at an average distance of $359.2\pm 2.5$~pc, placing them at the edge of the shell, about 66 pc from the shell center, and as close as $\sim$ 3 pc from the nearest section of the shell boundary (see also Figure \ref{fig:shell_w_stars}). Given that these six young stars are embedded within CG30 and are also located along the edge of the IVS, they provide a compelling evidence for the association between the IVS and CGs. This association further validates our adoption of $v_{\rm exp} = 12\pm 3\;\rm km\;s^{-1}$ as the IVS expansion velocity. 

\subsection{Origin of the IVS}

Alongside its relationship with the Gum Nebula, the origin of the IVS has been widely examined with various feedback mechanisms proposed to be involved, including stellar winds, HII region expansion, supernovae, and ultraviolet (UV) evaporation \citep[i.e. the ``rocket effect"; see e.g.][]{Sridharan1992, 2006Testori, 2019Cantat-Gaudin}. Such interplay of multiple feedback mechanisms is also shown in other large-scale structures, such as the Orion-Eridanus superbubble \citep{Foley_2023, Soler2018, Bally2008}.

In this paper, we concentrate on the impact of stellar winds, HII region expansion, and supernovae, three dominant mechanisms of momentum injection into the ISM on the scale of the IVS \citep[see e.g.][\S 4.2.1]{2018KimKimOstriker}. While UV evaporation plays a more significant role in dense, optically thick regions, such as cometary globules, less than 0.3\% of the gas on the IVS surface is optically thick (where column densities $N_{\rm H}\geq 10^{21}\rm \; cm^{-2}$ \citep[e.g.][]{Clark2014}). Moreover, due to their small angular sizes and high densities, the CGs cannot be resolved in the E24 map, where the rocket effect should be the most important. Therefore, we defer a detailed analysis of the rocket effect to future work.

\subsubsection{Contributions from Stellar Winds} \label{appendix: stellar_winds}

We first consider stellar winds. The dominant sources in the vicinity of the IVS that produce substantial stellar winds are $\zeta$ Puppis and $\gamma^2$ Velorum. As shown in Figure \ref{fig:shell_w_stars}, both star systems are located within our newly constrained 3D model of the IVS, making them likely candidates for powering the shell's expansion.

$\zeta$ Puppis is one of the brightest O supergiant stars in the solar vicinity that has been extensively studied \citep[e.g.][]{Sota2014, Howarth2019, Ramiaramanantsoa2022}. It is a hot and massive star with a mass of $25.3 \pm 5.3 \rm \, M_{\odot}$, an age of 2.2-3.6 Myr, and a peculiar velocity of $56.2 \pm 1.9 \rm \; km\;s^{-1}$ \citep{Howarth2019}, making it the main ionizing source of the Gum Nebula \citep{Chanot1983, Woermann2001}. The binary system $\gamma^2$ Velorum contains one Wolf-Rayet star (WC8) and one O5-O8 companion. Wolf-Rayet stars are known for their high mass-loss rates and strong stellar winds \citep[e.g.][]{DeMarco1999, DeMarco2000, Crowther2024}.

We adopt the formalism for the mechanical energy and momentum injection from an idealized stellar wind model, with mechanical wind luminosity given by:
\begin{equation}
    L_{\rm wind} = \dot{E}_{\rm wind} = \frac{1}{2} \dot{M}v_{\infty}^2,
\end{equation}
where $\dot{M}$ is the mass-loss rate, and $v_{\infty}$ is the terminal speed of the stellar wind \citep{2011Draine}.

The momentum injection rate is given by:
\begin{equation}
    \dot{p}_{\rm wind} = \dot{M}v_{\infty}.
\end{equation}
Combining the mass-loss and velocity data presented in Table~\ref{tab:stellar_wind} and integrating the mechanical luminosity over the each object's stellar age, we can compute the total energy contribution from stellar winds with 95\% C.I.:
\begin{equation}
    E_{\rm wind} = (6.0 \pm 0.8) \times 10^{50} \;\rm erg
\end{equation}
and the total momentum contribution from stellar winds: 
\begin{equation}
    p_{\rm wind} = (3.0 \pm 0.4) \times 10^{4} \; \rm M_{\odot}\;km\; s^{-1}.
\end{equation}

For the Wolf-Rayet (WR) component of the $\gamma2$ Velorum, we assume a typical WR phase duration of 0.5 Myr \citep{Maeder1991}. Given that WR stars experience significantly stronger winds than the cumulative winds from their O-type progenitors, we focus primarily on the WR-phase contribution for the WC8 component. A more detailed treatment incorporating the main-sequence phase mass loss could be explored in future work.

Recognizing that the young star cluster Vela OB2 also lies near the IVS, we apply the empirical prescription from \citet{Leitherer1992} to estimate the collective mass-loss rate and terminal velocities of the prominent B stars in the cluster, as no O-type stars have been identified \citep{deZeeuw1999}:
\begin{multline}\label{eqn:velaob2}
\begin{split}
    \rm log(\dot{M}) = -24.06 + 2.45\; \rm log(L) - 1.10\; log(M) + \\
   \rm 1.31\; log(T_{\rm eff}) +0.8\; log(Z);\\
    \rm log(v_{\infty}) = 1.23 - 0.30\; \rm log(L) + 0.55\; log(M) + \\
    \rm 0.64\; log(T_{\rm eff}) + 0.13\; log(Z)
\end{split}
\end{multline}
Our calculations show that the stellar wind momentum ($\simeq 1.2\times 10^{3}\;\rm M_{\odot}\cdot km\; s^{-1}$) and energy ($\simeq2.4\times 10^{49}\,\rm{erg}$) from Vela OB2 contribute less than 5\% of the total momentum and energy compared to the combined inputs from $\zeta$ Puppis and $\gamma^2$ Velorum. Therefore, in following analysis, we do not consider Vela OB2 as a significant feedback source.

 \citet{2006Testori} also analyzed the energy contributions from various sources. They found the dominant contribution arises from $\gamma^2$ Velorum of $(5-11) \times 10^{50} \, \rm{erg}$, which is slightly larger than our value, $\sim 2\times 10^{50}$ erg. We find that $\zeta$ Puppis can produce more stellar wind energy and momentum than $\gamma^2$ Velorum.

Recall that we obtain a total momentum of the IVS, $p \simeq 6 \times 10^{5}\;\rm M_{\odot}\cdot km\; s^{-1}$, from \S\ref{subsec:momentum_energy}. Comparing this with the total stellar wind momentum from $\gamma^2$ Velorum and $\zeta$ Puppis, $p_{\rm wind} \simeq 3\times 10^{4}\;\rm M_{\odot}\cdot km\; s^{-1}$, we find that the combined stellar wind momentum accounts for only $\sim 5\%$ of the total IVS momentum. Additionally, the complex star formation history in the IVS-Gum Nebula region introduces further uncertainty in assessing the cumulative contribution of stellar winds throughout the region's evolutionary history. In particular, the progenitors of the Vela Pulsar and pulsar RX J0720.4-3125 (as discussed further in \S\ref{subsubsec:pulsar}), along with other massive stars that eventually underwent supernova explosions, likely played a role in shaping the present-day morphology of the IVS. However, due to the significant uncertainties in the timing, spatial distribution, and progenitor properties of these supernova events, we refrain from a detailed analysis of their individual contributions. Nevertheless, the stellar winds alone are unlikely to account for the total expansion momentum of the IVS, indicating that there are other contributors to the total shell momentum.

Constraining the total effective energy contribution from stellar winds to the IVS is more challenging than estimating the momentum, as the energy depends heavily on the surrounding environment and can be lost through various processes, including radiative cooling, thermal conduction, and dust collisions \citep[e.g.][]{Rosen2014}. Empirically, the stellar wind energy conversion efficiency can range from 3.7\% to 38\% \citep{Rosen2014}, which means that the stellar wind energy can range from $2.2\times 10^{49}$ erg to $2.3\times 10^{50}$ erg of the IVS kinetic energy. Given the IVS kinetic energy $E_{k} \simeq 7.1\times 10^{49}$ erg, we find momentum is a relatively easier metric to constrain and a more meaningful metric for our study.

\begin{deluxetable*}{ccccccccc}
\rotate
\tablecaption{Physical properties of Objects in $\gamma2$ Vel and $\zeta$ Puppis  \label{tab:stellar_wind} }
\tablehead{
\colhead{Object} & \colhead{$\dot{M}$}                               & \colhead{$v_{\infty}$ }       & \colhead{$L_{\rm wind}$}                  &\colhead{t }               &\colhead{$E_{\rm wind}$  }              & \colhead{$p_{\rm wind}\,  $}                    & \colhead{$\rm{log}Q$ }  & \colhead{log$L_{\rm{bol}}$} \\
\colhead{}       & \colhead{$[\rm M_{\odot} \;\rm{yr^{-1}}] $}        & \colhead{[km$\;\rm{s^{-1}}$]} & \colhead{[erg$\;\rm{s^{-1}}$]}            &\colhead{[Myr]}            &\colhead{[erg]}                         & \colhead{$[\rm M_{\odot}\, \rm{km\;s^{-1}}]$}   & \colhead{[$\rm{s^{-1}}$]}    &\colhead{[$L_{\odot}$]}   }
\startdata
WC8 in $\gamma^{2}$ Vel     & $1.4\times 10^{-5}$ [1]                & $1500 \pm 20$ [1]             & $(1.0 \pm 0.1) \times 10^{37}$            & 0.5 [2]                   & $(1.6 \pm 0.1) \times 10^{50}$         & $(10.5 \pm 0.3)\times 10^{3}$                   & 48.81 - 48.75 [3]            &   5.31 [1]                         \\
O in $\gamma^{2}$ Vel       & $(1.78\pm 0.37) \times 10^{-7}$ [3]    & $2500 \pm 250$ [3]            & $(3.5^{+2.3}_{-1.7}) \times 10^{35}$      & $3.59 \pm 0.16$ [3]       & $(3.9^{+2.7}_{-1.9}) \times 10^{49}$   & $(1.6_{-0.7}^{+0.8})\times 10^{3}$              &  48.6-49 [3]                 &   5.56 [1]                         \\
$\zeta$ Puppis              & $(2.5\pm 0.2) \times 10^{-6}$ [4]      &  2250 [5]                     & $(4.0 \pm 0.6) \times 10^{36}$            & $3.2^{+0.4}_{-0.2}$ [6]   & $(4.0\pm 0.8) \times 10^{50}$          & $(18.0_{-3.4}^{+3.7})\times 10^{3}$             &  49.70 [7]                   &   5.60 [4]                         \\
\enddata
\tablecomments{$\dot{M}$: mass-loss rate; $v_{\infty}$: terminal speed of stellar wind; $L_{\rm{wind}}$: stellar wind luminosity; t: stellar age; $E_{\rm wind}$: total stellar wind energy (integrated over the stellar age); $p_{\rm{wind}}$: stellar wind momentum; Q: ionizing ($h\nu>13.6 \rm \;eV$) photon emission rate; $L_{\rm{bol}}$: bolometric luminosity. 1$\sigma$ uncertainties taken from the literature are shown for the cited values of ($\dot{M}$, $v_{\infty}$, t) while 95\% C.I. are provided for the calculated values ($L_{\rm{wind}}$, $E_{\rm wind}$, $p_{\rm{wind}}$). References -- 
[1]: \citet{Crowther2024}, [2]: \citet{Maeder1991}, [3]: \citet{DeMarco2000}, [4]: \citet{Howarth2019}, [5]: \citet{Puls2006}, [6]: \citet{Bouret2012}, [7]: \citet{2011Draine}}
\end{deluxetable*}

\subsubsection{Contributions from HII Region Expansion \label{subsubsec:HII region}} 

The hot stars within the IVS-Gum Nebula system, including $\zeta$ Puppis and $\gamma^2$ Velorum, emit ionizing radiation capable of powering HII regions and heating the surrounding gas to temperatures on the order of $10^{4}\,\rm{K}$. A simple order-of-magnitude estimate for the Lyman continuum (LyC) ionizing photon luminosity of an HII region with the size of the IVS is given by: $Q_{\rm{LyC}} = 4\pi/3\:r_{\rm{IVS}}^{3}\:n_{\rm{H}}^{2}\:\alpha_{B}\approx4.5\times10^{48}\;\rm s^{-1}$ \citep[]{2011Draine}, where $\alpha_{B}$ is the case-B recombination coefficient for hydrogen, $r_{\rm IVS}= 70\;\rm pc$ is the IVS radius, and $n_{\rm H} = 0.67\;\rm cm^{-3}$ is the adopted ambient density (calculated in \S\ref{subsec: density_ambient}). Given that the H$\alpha$ morphology and the projected 3D dust morphology are roughly co-spatial on the plane of the sky (see Figure \ref{fig: Halpha_w_IVS}), we use the same radius for the HII region as the 3D-dust-traced IVS. The calculated $Q_{\rm{LyC}}$ is lower than the LyC photon emission rate of the Wolf-Rayet star in the $\gamma2$ Velorum binary alone \citep[see Table \ref{tab:stellar_wind}]{Crowther2024}, underlining the importance of also considering the feedback from HII region expansion. Additionally, the thermal energy of the HII region, assuming an ideal gas, is given by $E_{\rm{therm}} = 4\pi/3\:r_{\rm{IVS}}^{3}\:n_{\rm{H}}\:k_{B}\:T = 1\times10^{50}\; \rm erg$ where we adopt a temperature of $T\approx 11,300\;\rm K$ for the Gum nebula \citep{Reynolds1976I}. This value is comparable to IVS expansion kinetic energy. These estimates of ionizing photon luminosity and thermal energy motivate us to conduct a more detailed investigation into the role of photoionization and HII region expansion in the IVS-Gum Nebula system, with an emphasis on radiative and thermal momentum injection.

The contribution of thermal pressure to the dynamics of HII regions remains an active research topic in both simulations and observations \citep[e.g.][]{Jeffreson2021, Olivier2021}. Here, we estimate the thermal momentum injection using the simple relation adapted from \citet[][see their Eqn. 3]{Jeffreson2021}:
\begin{equation}
    dp_{\rm{therm}}/dt = A_{\rm{IVS}}P_{\rm{therm}},
\end{equation}
where $A_{\rm{IVS}}\approx2\pi r_{\rm{IVS}}^{2}$ is the surface area of a blister-type HII region, and $P_{\rm{therm}} \sim 2\rho_{\rm{IVS}} c_{\rm{IVS}}^{2}$ represents the thermal pressure, with $\rho_{\rm{IVS}}$ being the density of the ionized gas interior to the swept-up shell\footnote{Here we adopt the present-day interior number density $0.17\;\rm cm^{-3}$ for the calculation following Table \ref{tab:property}, with the assumption that the interior is fully ionized.} and $c_{\rm{IVS}}= \sqrt{2kT/m_{\rm H}}$ the speed of sound\footnote{Here we use $T\approx 11,300\;\rm K$ as mentioned in the text}. This yields an instantaneous momentum injection rate of $4.9\times10^{4}\;\rm{M_{\odot}}\;\rm{km\;s^{-1}\;Myr^{-1}}$. Assuming a constant rate, the upper limit for the total thermal pressure-driven momentum injection over the system's lifetime is $\sim1.5\times10^{5}\;\rm{M_{\odot}}\;\rm{km\;s^{-1}}$, comparable to the IVS mechanical momentum. However, we emphasize that this estimate carries significant uncertainties, and a more detailed analysis is reserved for future studies.

The momentum imparted by ionizing photons is given by: $dp_{\rm{phot}}/dt = f_{\rm{trap}}L_{\rm{bol}}/c$, where $L_{\rm{bol}}$ is the bolometric luminosity of the primary ionizing sources, and $f_{\rm{trap}}$ accounts for the enhancement of radiation pressure from energy trapping within the shell \citep{Krumholz2009}. Integrating over the lifetimes of $\zeta$ Puppis and $\gamma^2$ Velorum, we obtain the total radiative momentum contribution:  

\begin{equation}
     p_{\rm{phot}} = \sum_{\gamma\rm{2Vel},\zeta \rm{Pup}}f_{\rm{trap}}L/c \cdot t = f_{\rm{trap}}\cdot 2.8\times 10^{4}\; \rm M_{\odot}\;km\;s^{-1}.
\end{equation}

Theoretical constraints on radiation pressure are relatively well-established \citep[as shown, e.g.][]{Krumholz2009, Rahner2017}, whereas $f_{\rm{trap}}$ remains highly environment-dependent, influenced by stellar winds, infrared photons, and Ly$\alpha$ photons. Based on observational data from various HII regions, \citet{Olivier2021} found a median $f_{\rm{trap}}\approx8$; however these regions are orders of magnitude smaller than the Gum-IVS system (with radii of $< 0.5 \;\rm pc$) making $f_{\rm{trap}}\approx8$ a likely upper limit. Using recent results from \citet[submitted][private communication]{McCallum2025}, who modeled H$\alpha$ emissivity across the Milky Way by simulating Lyman continuum photon propagation through the solar neighborhood --- based on the same \citet{Edenhofer2024} 3D dust map adopted here --- we estimate that only $\sim 21\%$ of the LyC photons produced by $\gamma^2$ Velorum and $\zeta$ Puppis are absorbed within the IVS. These results indicate that the IVS is a relatively ``leaky” structure, \textit{optically thin} to LyC photons. Consequently, assuming $f_{\rm trap}$ in the order of $1-10$ the total momentum contribution from the radiation pressure is estimated to be on order of $10^{4}-10^{5}\; \rm M_{\odot}\;km\;s^{-1}$.

The combined momentum imparted by the HII region from thermal expansion and radiation is of the same order of magnitude as the IVS mechanical momentum. However, given the large uncertainties in thermal momentum estimates and the presence of fast-moving stellar objects (e.g., $\zeta$ Puppis) in the region, it remains inconclusive whether these two mechanisms are sufficient to heat the ISM, ionize the gas, and sustain the current expansion of the IVS simultaneously, as discussed in \citet{Reynolds1976}. While our analysis suggests that stellar winds are subdominant, multiple feedback mechanisms are likely acting in concert. In particular, the role of supernova should also be considered alongside HII region expansion to fully characterize the energy and momentum budget of the system.

\subsubsection{Contribution from Supernovae}

In this section we focus on supernovae as another potential mechanism to help explain the observed expansion and morphology of the IVS. Following \citet{Bialy2021, Foley_2023, Zucker2022}, we estimate the number of supernovae required to produce the shell and its expansion velocity. To start, we compare the total shell momentum to the typical momentum injected per supernova explosion. The typical momentum imparted to the ISM by a supernova explosion is $\hat{p} = (2 - 5)\times 10^5 M_\odot\; \rm{km \;s^{-1}}$~\citep{2019El-Badry}. We assume that $\hat{p}$ can be uniformly distributed in this range, and we sample the momentum of the IVS $p$ from the distribution obtained in \S\ref{subsec:momentum_energy}. This lets us sample the number of supernovae required to drive the IVS, with the median value and 95\% C.I. given by
\begin{equation}
    N_{\rm SN} = \frac{p}{\hat{p}} = 1.7_{-1.1}^{+2.2},
\end{equation}
where we denote the number of supernovae by $N_{\rm SN}$. 

\begin{figure}
\includegraphics[width=.5\textwidth]{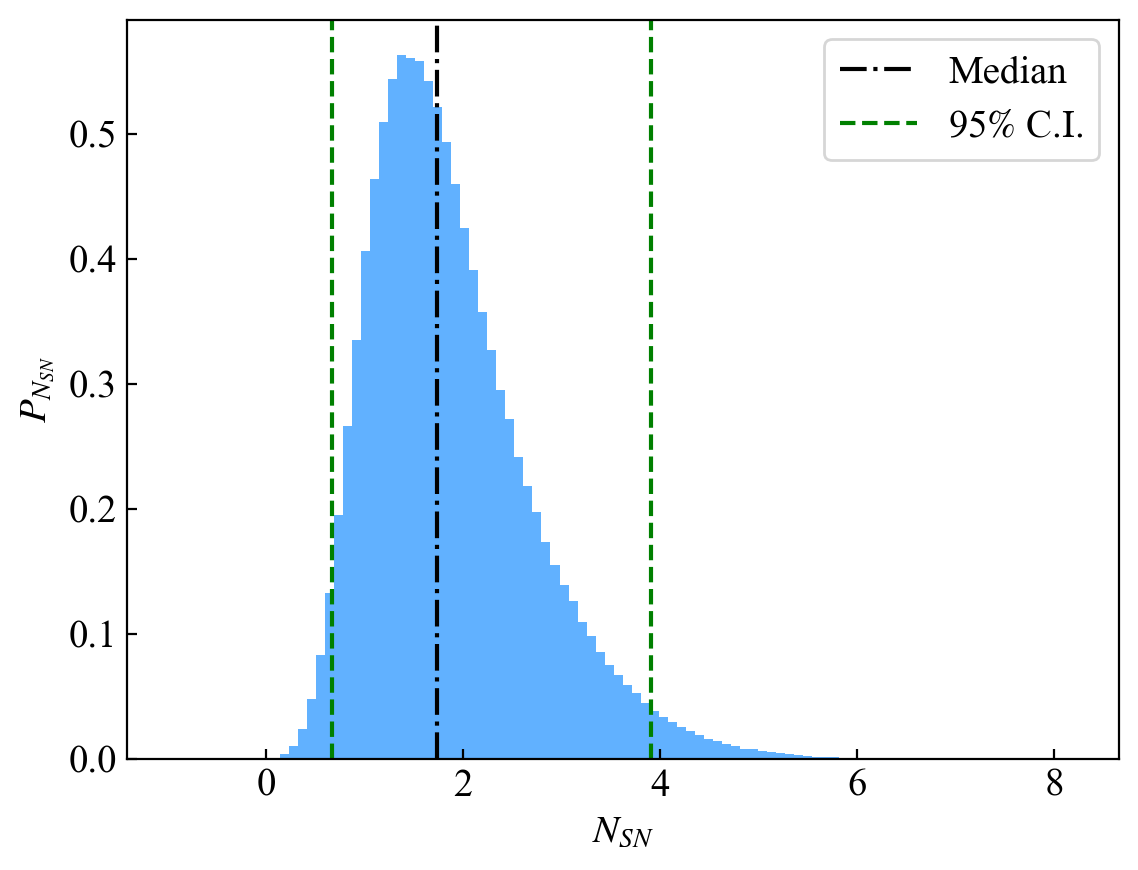}
    \caption{Probability density distribution of the number of supernovae required to account for the total momentum of the IVS.}
    \label{fig:n_SN}
\end{figure}

The probability distribution of the number of supernovae is shown in Figure \ref{fig:n_SN}. 
We find that one or two supernovae are theoretically enough to power the expansion, making this a likely energy source for the IVS. 

The non-zero number of supernovae computed above leads us to delve deeper --- to look for evidence of past supernova activity and discuss the potential sources. One obvious supernova remnant is the Vela Pulsar currently located inside the IVS as presented in Figure \ref{fig:shell_w_stars}. However, this pulsar has an age of around 20,000 years old \citep{Becker2002}, significantly younger than the estimated age of the IVS. Hence, this pulsar has a negligible impact on the dynamics and evolution of the IVS. In the following subsections, we mainly focus on finding the supernova that likely cause the formation of the IVS. 

\subsubsection{Dynamical Age of the IVS and Traceback Analysis}\label{subsubsec:IVS age}
Assuming that the IVS could be a potential supernova remnant, we begin by estimating the dynamical age of the shell. 
Given the current size and expansion rate of the shell, the expected dynamical age of the shell is estimated as 
 \begin{equation}
     t_{\rm dyn} =\eta \frac{r_{\rm{shell}}}{v_{\rm{exp}}},
 \end{equation} 
 where $r_{\rm{shell}}$ is the shell radius, $v_{\rm{exp}}$ is the expansion velocity of the shell, and $\eta$ is the expansion parameter that depends on what stage the remnant is in \citep{2011Draine} and whether it is powered by a single supernovae or continuous supernovae \citep[see e.g. Eqn. 2 and references therein from][]{ Bialy2021}. For a single supernova, $\eta = 0.25-0.3$~\citep{McCray1987, Bialy2021}. As before, we sample the expansion velocities from a Gaussian distribution with $N(\mu=12, \sigma=3)\rm \; km \; s^{-1}$~\citep{Rajagopal1998} and radius re-sampled from a flattened array where each ray was sampled from its corresponding Gaussian distribution of $N(\mu=\rm{\mu_{ray}}, \sigma=\rm{\sigma_{ray}})$. The resulting median and 95\% C.I. of the dynamical age is 
 \begin{equation}
    t_{\rm dyn}~=~1.6_{-0.6}^{+1.5} \; \rm{Myr}.
 \end{equation}

This value is comparable to the timeline proposed by \citet{Woermann2001}, who suggested that the Gum Nebula, together with a neutral shell inside the region, was likely produced by a supernova event that occurred 0.42 to 1.4 Myr ago. Compared to the estimate by \citet{2019Cantat-Gaudin} who proposed that a supernova explosion occurred 10 to 20 Myr ago based on the age of the star cluster Vela OB2, our estimation suggests a much more recent event. In the following analysis, we investigate potential supernova sources within the vicinity of the IVS over the past $\sim 3\rm\; Myr$.

To determine the trajectories of potential sources of supernova feedback, we employ a traceback analysis following the method used by \citet{Zucker2022, Swiggum2024, Edenhofer2024_cham}, where we assume the peculiar motion of the Sun to be (11.1, 12.24, 7.25) $\rm km\;s^{-1}$ \citep{Schonrich2010}. By drawing 10,000 samples from the normal distribution of each object's 3D position and 3D velocity, as listed in Table \ref{tab: trace_back_astrometry}, we integrate their possible range of trajectories backward in time over a period of 3 Myr with a step size of 0.01 Myr. To do so, we leverage the \texttt{galpy} Python package and its standard \texttt{MWPotential2014} \citep{Bovy2015}. This analysis enables us to pinpoint the relative positions of nearby objects over time and examine their potential interactions with the IVS. Figure \ref{fig:stars_clusters_path} presents an overview of the locations of key objects at different time snapshots relative to the IVS center in the LSR frame. 

Following the supernova explosion, we estimate the timescale for shell formation \citep[after the end of the Sedov-Taylor stage;][]{Taylor_1950} from \citet{KimOstriker2015}. Using the prescription from \citet{KimOstriker2015} (Eqns. 7, 8), the time of shell formation and the outer shell radius at the time of formation is given as: 
\begin{align}    
   t_{\rm{sf}} &= 4.4\times 10^{4}\;{\rm{yr}}\; E_{51}^{0.22}\; n_{0}^{-0.55} = 5.5\times 10^{4}\;{\rm{yr}} \\
   r_{\rm{sf}} &= 22.6\;{\rm{pc}}\;E_{51}^{0.29}\;n_{0}^{-0.42} = 26.7\;{\rm{pc}} 
\end{align}
where the value of $n_{0}$ is taken from Table \ref{tab:property}, assumed to be our predicted ambient density. After formation, we assume the IVS expands with its radius scaling as $R \propto t^{\eta}$ \citep[c.f.][]{Bialy2021, Kim2005}. 

\begin{deluxetable*}{ccccccccc}
\tablecaption{Astrometric Data for Each Object Applied in the Traceback Analysis\label{tab: trace_back_astrometry}}
\colnumbers
\tablehead{\colhead{Object} & \colhead{$\alpha$ } & \colhead{$\delta$} &  \colhead{$\varpi$} & \colhead{(X, Y, Z)} & \colhead{$\mu_{\alpha*}$ } & \colhead{$\mu_{\delta}$} &\colhead{RV} & \colhead{(U, V, W)} \\
\colhead{} & \colhead{[$^\circ$]} & \colhead{[$^\circ$]} & \colhead{[mas]} & \colhead{[pc]} & \colhead{[mas\,yr$^{-1}$]} & \colhead{[mas\,yr$^{-1}$]} & \colhead{[km\,s$^{-1}$]} & \colhead{[km\,s$^{-1}$]} }
\startdata
$\zeta$ Puppis \tablenotemark{\tiny a} & 120.9 & -40.0   & $3.01 \pm 0.10$ &  (-80, -321, -27)  & $-29.71 \pm 0.08$ & $16.68 \pm 0.09$ & $-25.4 \pm 2.1$ &  (-39.0, 38.0, -23.7)\\
Trumpler 10 \tablenotemark{\tiny b} & 131.9 & -42.5   & $2.29 \pm 0.07$ & (-54, -432, 5)  &$-12.38 \pm 0.30$ & $6.57 \pm 0.25$ &  $20.00 \pm 9.78$ & (-28.9, -16.6, -11.2) \\ 
RX J0720.4-3125 \tablenotemark{\tiny c} & 110.1 & -31.4   & $3.6 \pm 1.6$  & (-120, -247, -39) & $-92.8 \pm 1.2$ & $55.3 \pm 1.3$ &  --  & -- \\
\enddata
\tablecomments{(1) Name of the object. (2 - 4) Right Ascension $\alpha$, declination $\delta$, and parallax $\varpi$ of the objects. (5) Position of the object in Heliocentric Galactic Cartesian Coordinate. (6-8) Kinematics of the object, including proper motion $\mu_{\alpha*}$, $\mu_{\delta}$ and radial velocity. (9) Heliocentric Galactic Cartesian velocities of the object. All uncertainties correspond to $1\sigma$. Machine-readable version is available online at the Harvard Dataverse \url{https://doi.org/10.7910/DVN/QF2VGG}.}

\tablenotetext{a}{Data obtained from Hipparcos catalogue \citet{1997HIPPARCOScatalog}.}
\tablenotetext{b}{Cluster astrometry obtained from catalogue produced by \citet{Hunt2023} whose work is based on Gaia DR3 \citep{Gaia2021} measurements.}
\tablenotetext{c}{ $\alpha$ and $\delta$ are adopted from \citet{Kaplan2007} whose work is based on Hubble space telescope. The rest of astrometry data are adopted from the measurement of \citet{Eisenbeiss2011} who observed the object with VLT-FORS1 at the ESO-Paranal.}
\end{deluxetable*}

\subsubsection{Potential SNe Source: the Companion of $\zeta$ Puppis}
$\zeta$ Puppis, a prominent runaway O-type star located at the front edge of the IVS (green arrow in Figure \ref{fig:stars_clusters_path}) has long drawn attention for its potential impact on the surrounding ISM. Runaway stars like $\zeta$ Puppis can be produced through supernova explosions in massive binary systems or through gravitational interactions within star clusters \citep{Blaauw1961, Poveda1967, Hoogerwerf2001}. In the case of $\zeta$ Puppis, its presence has been linked to recent supernova activity~\citep{Woermann2001}, with studies suggesting its binary progenitor likely originated from the Trumpler 10 star cluster~\citep{Schilbach2008, Hoogerwerf2001}. 

To confirm this association, we apply the traceback analysis to $\zeta$ Puppis and Trumpler 10. In our analysis, the average smallest distance between the two (with 95\% C.I.) is $9.5_{-8.2}^{+15.0}$ pc, at $2.0_{-1.0}^{+0.7}$ Myr ago, as shown in Figure \ref{fig:stars_clusters_path} at the -2 Myr subplot. The absolute minimum approach among all 10,000 samples is 0.17~pc, falling well within the Trumpler 10 cluster radius of 1.38 pc, which is the distance where the overall cluster has the best contrast to field stars \citep[see \S 3.3 of][]{Hunt2023}. This finding suggests that $\zeta$ Puppis could indeed have originated in Trumpler 10 in the past. Notably, this time of closest approach aligns with the 95\% C.I. of $t_{\rm dyn}$. However, the location of the closest approach is far outside the IVS shell boundary, approximately $120-160$~pc from its current geometric center, as illustrated in Figure~\ref{fig:stars_clusters_path}. 

Assuming that the center of the IVS moves with the Local Standard of Rest (LSR), $\zeta$ Puppis passed within 50 pc of the geometric center around 0.17 Myr ago. Trumpler 10, on the other hand, has always been outside of the IVS, except in the unlikely scenario where the shell moved a significant distance ($> 100$ pc) with respect to the LSR since its formation. As such, although our analysis supports past studies about $\zeta$ Puppis originating from a Trumpler 10 supernova \citep{Hoogerwerf2001, Schilbach2008}, we challenge the scenario where its former binary companion was the supernova driving the expansion of the IVS. Additionally, we acknowledge that the dynamical interactions between a stellar binary and its parent cluster can be more complex than our traceback analysis can fully capture --- for instance, around $20\%-30\%$ of supernova progenitors are found outside their original stellar groups \citep{Mason1998, Maiz-Apellaniz2004,Tetzlaff2013}. In our case, this means that the companion of the $\zeta$ Puppis could have gone supernova after the binary left the main star cluster, opening avenues for future investigation. We note that \citet{SridharanThesis1992} invoked a progenitor massive binary system inside the current IVS with $\zeta$ Puppis as the less massive component to explain the system of cometary globlues.

\subsubsection{Potential SNe Source: RX J0720.4-3125}\label{subsubsec:pulsar}
Another supernova candidate in the IVS region is the pulsar RX J0720.4-3125, a young isolated neutron star with a relatively precise parallax measurement \citep{Motch2003, Kaplan2007}. The radial velocity of RX J0720.4-3125 remains unconstrained, leaving its origin under debate. \citet{Kaplan2007, Tetzlaff2011, Tetzlaff2013} suggested that RX J0720.4-3125 also may have originated in the Trumpler 10 cluster based on parallax and proper motion data, considering a radial velocity broadly sampled across the range $[-500, 500]\; \rm km\; s^{-1}$.

To investigate this candidate, we perform a traceback of the pulsar, sampling the radial velocity uniformly over the interval $[-500, 500]\; \rm km\; s^{-1}$ following \citet{Tetzlaff2011}. We find that fewer than $5\%$ of the sampled trajectories pass within 70 pc of the shell center and no samples pass nearer than 60 pc to the shell center. Therefore, it is unlikely that this neutron star is tied to the formation of the IVS given its present day morphology. Next, we compute the minimal approach between RX J0720.4-3125 and Trumpler 10 to be about 17 pc, finding the meeting time ($0.57_{-0.51}^{+1.79}$ Myr ago, with 95\% C.I.) and the corresponding pulsar radial velocity $-98_{-119}^{+134}\;\rm km\;s^{-1}$ in strong agreement with the findings of \citet{Tetzlaff2011}. Therefore, RX J0720.4-3125 could have originated in Trumpler 10 and led to a separate nearby supernovae explosion (just exterior to the IVS) in the past few million years, contributing to the overall high supernova rate in this part of the Galaxy (see \S \ref{subsubsec:nearby clusters}).

\subsubsection{Other Nearby Prominent Star Clusters}\label{subsubsec:nearby clusters}

Although we are not able to definitely trace the supernova back to the former companion of $\zeta$ Puppis or the neutron star RX J0720.4-3125, other nearby young, massive clusters may have produced supernovae over the past few million years. A recent study by \citet{Swiggum2024} revealed that the majority of young star clusters (younger than 70 Myr) in the Vela region are primarily associated with either the Collinder 135 (Cr135) or Messier 6 (M6) cluster families, with the clusters in each group sharing a common formation history. The Cr135 family, with a median age of 22 Myr, includes 39 clusters and likely generated approximately 60 supernovae in the past 40 Myr. The M6 family, comprising 34 clusters and including Trumpler 10, has a median age of 32 Myr and likely produced around 120 supernovae in the past $\sim 60$ Myrs. 

\citet{Swiggum2024} also identified a smaller family of eight clusters that are located within the IVS region, with a median age of 9 Myr. The family includes the $\gamma$ Vel cluster (Pozzo 1), associated with $\gamma^2$ Vel \citep{2000Pozzo, 2009Jeffries}, suggesting that the family might represent the lower-mass component of the Vela OB2 association \citep{2000Pozzo, 2019Cantat-Gaudin}. Hereafter, we refer to this cluster family as the $\gamma$ Vel family. The complex star formation history in this region motivates our exploration of young clusters that may be relevant to the formation of the IVS.

\begin{figure*}\includegraphics[width=1\textwidth]{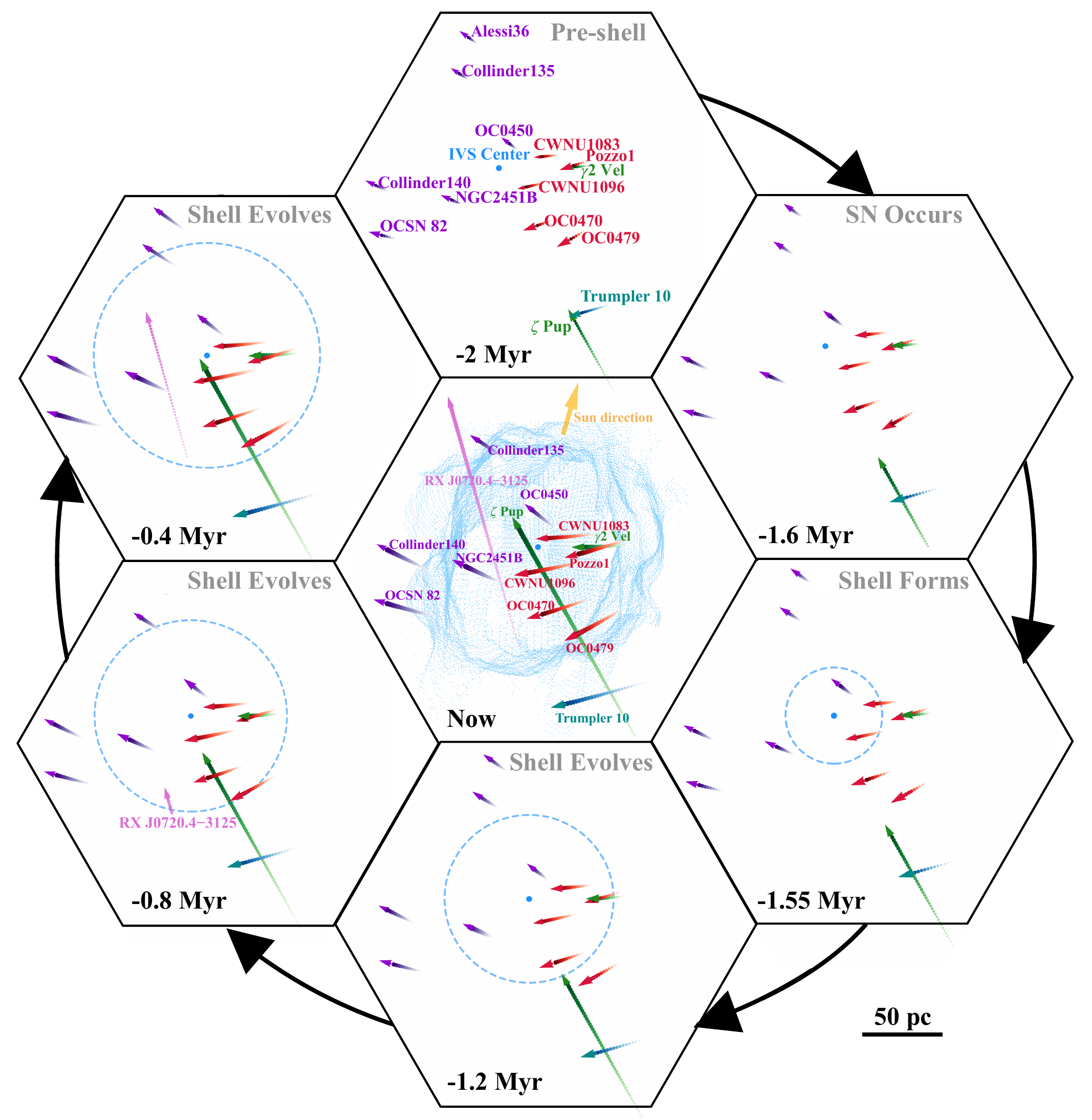}
    \caption{A series of top-down views in the x-y plane, illustrating the relative locations of key objects in the vicinity of the IVS (the fiducial model of the IVS peak, colored in blue) in the LSR frame. Before the explosion of the supernova 1.6 Myr ago, the shell is marked with its center location. After its formation at -1.55 Myr (in the third panel), we keep the shell outline to guide the eye, while marking the shell expansion following $R\propto t^{\eta}$, as discussed in \S\ref{subsubsec:IVS age}. For pulsar RX J0720.4-3125 which has no radial velocity constraints, we plot one scenario where it traverses the IVS; details are discussed in \S\ref{subsubsec:pulsar}. The paths of two sets of nearby massive clusters that could have hosted supernovae in the past 3 Myr are shown in purple and red (see discussion in \S\ref{subsubsec:nearby clusters}). The interactive version of this figure can be found \href{https://annie-bore-gao.github.io/Images/gao24_fig8_interactive.html}{here}. Users can also switch to the view in the $\gamma$ Vel cluster family frame (see \S\ref{subsubsec:nearby clusters}) using the button within the interactive figure.}
    \label{fig:stars_clusters_path}
\end{figure*}

If a stellar cluster hosted a progenitor capable of powering the shell, it should trace back to the center of the shell around one dynamical age ago in time. However, accurately tracing these clusters back to the shell's center requires knowledge of the shell's rest frame, which is not well-constrained. Therefore, we conduct two selections to identify possible progenitor clusters: a general selection applying the traceback analysis in the LSR frame (as shown in the static version of Figure \ref{fig:stars_clusters_path}), and a second selection in the $\gamma$ Vel family's rest frame, given that the $\gamma$ Vel family is currently situated within the shell's interior. 

First, for the general selection, we leverage the cluster catalog from \citet{Swiggum2024}, selecting clusters that have likely produced supernovae in the past few million years. We require these clusters to lie within a cubic volume of $(200\times 200 \times 200)\rm\; pc^{3}$ volume centered on the present-day shell center, be younger than $\leq$~40 Myr, and have a sufficient number of members ($\geq$~150). Then in the Galactic LSR frame, we further select the ones whose paths have intersected the present-day shell boundary and are moving away, rather than towards, the shell's center. This threshold is relatively conservative, since the IVS is expanding over time. After filtering, we are left with 6 clusters: Alessi 36, Collinder 135, Collinder 140, NGC2451B, and OC 0450, OCSN 82.

We augment this general cluster selection with a targeted investigation of a subset of the $\gamma$ Vel family: Pozzo 1, OC 0470, CWNU 1083, CWNU 1096, and OC 0479. These clusters present a plausible origin for the IVS for several reasons. First, they have ages of 5 to 9 Myr and their past trajectories converge to within 20-40 pc of each other around 11 Myr ago \citep[private communication]{Swiggum2024}. Second, they are all currently located within the IVS outer boundary. Applying the traceback analysis to these five clusters in their local frame (computed by averaging their 3D positions and 3D velocities), we find that these clusters move outward relative to the IVS center, with their traces remaining within the IVS radius. The interactive figure in the $\gamma$ Vel frame is available at: \url{https://annie-bore-gao.github.io/Images/gao24_fig8_interactive.html}. This suggests that, if the IVS is drifting along with the average movement of the $\gamma$ Vel cluster family (rather than the LSR), the five clusters could have feasibly contributed to the formation of the IVS. 

Combining the results of our general cluster selection with our targeted $\gamma$ Vel selection, we have a total of 11 clusters that are potentially associated with the formation of the IVS. \citet{Swiggum2024} estimated the expected number of supernova events for each star cluster in the solar neighborhood by fitting and sampling the cluster initial mass functions (IMFs), and counting the number of stars that are both more massive than $8\;\rm M_{\odot}$ and more massive than the current most massive star in the cluster. 

In Table \ref{tab:star_cluster}, we present the expected number of supernova events for each of the selected clusters over the past 3 Myr, $\mathbb{E}[\rm SN]$, as computed by \citet{Swiggum2024} [private communication]. To estimate the overall expected number of supernovae occurring within the IVS region in the past 3 Myr, we sum the $\mathbb{E}$[$\rm SN$] across the clusters. The first cluster selection criterion yields an estimated 1.16 supernovae while the second yields 1.21 supernovae. Both results align closely with our independent estimate of the number of supernovae needed to power the IVS (one to two supernovae). 

Our second selection can hint towards the following scenario: the $\gamma$ Vel family began forming in a molecular cloud around 9 Myr ago. After disruption of the parent molecular by feedback processes such as UV radiation and stellar winds, a supernova produced by the family swept up the remaining gas from the parent cloud, contributing to the formation of the IVS, where the family is now located. The combined stellar mass of these clusters is $\sim 10^{3}~\text{M}_{\odot}$ \citep{Swiggum2024}. Assuming $1\% - 3\%$ of the parent cloud formed these five clusters \citep{Evans1991, Kennicutt1998}, the remaining gas would be in the range of $3\times10^{4} - 10\times10^4~\text{M}_{\odot}$, in rough agreement with derived IVS mass (Table \ref{tab:property}). This scenario implies that as the IVS expanded, it drifted together with the $\gamma$ Vel family with a velocity of $(-0.98)\;\rm km\;s^{-1}$ in the LSR frame. On the other hand, even if the IVS is not drifting in the rest frame of the $\gamma$ Vel family, the clusters from our general selection criterion can cumulatively produce roughly one supernova. Hence, the formation scenario is consistent, independent of the exact cluster selection. 

We compare our estimated IVS origin to that proposed by \citet[hereafter, TCG19]{2019Cantat-Gaudin} based on Gaia DR2 data. According to their model, a set of clusters formed around 30 Myr ago, followed by the onset of supernova events approximately 10 Myr later. These supernova events created the IVS as well as initiating its expansion, inferring a shell age of about 20 Myr. As the IVS expanded, it triggered the formation of younger clusters, including Vela OB2. TCG19 identified eleven main sub-groups in Vela OB2 that appear to be expanding, which suggests that the expanding shell might have influenced the sequential formation of these younger stellar populations. This scenario aligns in part with our findings regarding the $\gamma$ Vel family, as we also observe evidence of younger clusters near the IVS boundary. Notably, the location and morphology of the regions A, C, G in TCG19 appear to correspond to Pozzo 1, OC 0479, and CWNU 1083 respectively, in our study. 

However, our analysis suggests a different timeline for the IVS. While TCG19 associated the IVS formation with the 30 Myr clusters (e.g. Collinder 135, NGC 2451 B, Trumpler 10), our findings imply a much younger shell. We suggest that the supernovae responsible creating the IVS likely occurred well after their peak supernova activity period, while earlier supernova may have partially evacuated the region. Notably, older clusters in this region have been associated with kiloparsec-scale dust activities \citep{Swiggum2024}, including the GSH 238+00+09 supershell \citep{Heiles1998}, which formed 30 Myr ago and surround the IVS. Alternatively, under our second selection criterion, these older clusters may not have directly caused the IVS formation. Instead, the younger clusters, which TCG19 associate with Vela OB2, might be the drivers of the IVS formation through recent supernova activity, rather than being products of the shell-triggered star formation. In this view, the younger stellar population would be directly responsible for the IVS, indicating a more recent star formation episode as the primary influence. 

\begin{deluxetable*}{ccccccc}
\tablecaption{Selected Star Clusters that Could Contribute to the IVS Expansion in the Past 3 Myr  \label{tab:star_cluster} }
\colnumbers
\tablehead{\colhead{Cluster Name } & \colhead{Family} & \colhead{$N_{\rm{star}}$ \tablenotemark{\tiny a}} & \colhead{Age [Myr] \tablenotemark{\tiny a} } &  \colhead{$\mathbb{E}$ [$\rm SN$] \tablenotemark{\tiny b} } & \colhead{(X, Y, Z) [pc] \tablenotemark{\tiny b} } & \colhead{(U, V, W) [$\rm km\;s^{-1}$] \tablenotemark{\tiny b} }}
\startdata
Alessi 36 & Cr 135 & 198 & 28.9  & 0.11 & (-99, -249, -64)  &  (-17.4, -7.5, -11.0)\\ 
Collinder 135 & Cr 135 & 209 & 29.8   & 0.11 & (-105, -271, -57)  & (-18.0, -7.5, -11.8) \\
Collinder 140 & Cr 135 & 203 & 17.3 & 0.17 & (-158, -338, -52) & (-19.8, -7.7, -11.2) \\
NGC 2451B & Cr 135 & 615 & 27.6 & 0.42 & (-111,-348,-44) & (-18.8, -8.4, -12.0)\\
OC 0450 & Cr 135  & 398 & 21.7 & 0.23 & ( -72, -312, -44) & (-16.7,  -7.6, -13.1)\\
OCSN 82 & Cr 135 & 157 & 23.6 & 0.12 & (-158, -370, -78) & (-21.6, -9.1, -7.5) \\
Pozzo 1 & $\gamma$ Vel & 365 & 9.4 & 0.44 & (-43,-341,-46) & (-20.9, -14.8,  -2.9)\\
OC 0470 & $\gamma$ Vel & 415 & 7.0 & 0.41 & (-65, -378, -69) & (-21.3, -15.2,  -3.0)\\
CWNU 1083 & $\gamma$ Vel & 100 & 7.7 & 0.08 & (-64, -332, -11) & (-22.1, -12.8, -1.5) \\
CWNU 1096 & $\gamma$ Vel & 23 & 4.6 & 0.04 & (-77, -353, 0.2) & (-23.7, 14.4, 0.4) \\
OC 0479 & $\gamma$ Vel & 92 & 7.2 & 0.24 & (-45, -390, -71) & (-21.1, -17.4, -4.8)\\
\enddata
\tablecomments{The two sets of clusters selected in the LSR frame and the $\gamma$ Vel family frame. Note that while Trumpler 10 is relevant to the broader IVS context, it does not meet the selection criteria in either frame and is therefore not included in this table. (1) Cluster name. (2) Cluster family. (3) Number of member stars. (4) Cluster age. (5) Expected number of supernova events over the past 3 Myr. (6) Current Heliocentric Galactic Cartesian Coordinate of the cluster. (7) Galactic Cartesian velocity. Machine-readable version is available online at the Harvard Dataverse \url{https://doi.org/10.7910/DVN/QF2VGG}.}
\tablenotetext{a}{Data adopted from \citet{Hunt2023}.}
\tablenotetext{b}{Data come from \citet{Swiggum2024} and used for traceback analysis.}
\end{deluxetable*}

\section{Conclusion} \label{sec: conclusion}

In this work, we derived the 3D geometry of the IRAS Vela Shell (IVS), computed its physical properties, and explored its formation history. Our main results can be summarized as follows:

\begin{itemize}
\item We demonstrated that the IVS is a prominent cavity associated with the Gum Nebula and found that the shell traces the morphological structure of the region's $\rm H\alpha$ emission on the plane of the sky. The shell center is located at a distance of 353 pc from the Sun and has a typical diameter of 138 pc. Using its associated young stars detected in Gaia, we place at least one cometary globule (CG30) associated with the Gum Nebula region on the surface of the shell. 

\item The IVS has a mass of $5.1_{-2.4}^{+2.4}\times 10^{4}\;\rm M_{\odot}$, momentum of $6.0_{-3.4}^{+4.7}\times 10^{5}\;\rm M_{\odot}\;km\; s^{-1}$, and energy of $7.1_{-5.4}^{+10.8}\times 10^{49}\rm erg$. The predicted ambient density prior to shell formation is estimated to be $0.67\pm 0.26\; \rm cm^{-3}$, approximately four times denser than the present-day shell interior density. The tenuous nature of this predicted ambient medium suggests that this region was likely shaped by earlier generations of feedback activity. 

\item Based on the total momentum of the IVS, we quantified the contributions from stellar winds, HII region expansion, and supernovae. Our analysis indicates that the shell is driven by the combined effects of photo-ionization (HII expansion) and supernovae feedback, with stellar winds contribute only about 5\% of the total momentum. We confirmed the significance of HII region expansion and found that the massive stars inside the IVS, $\gamma2$ Velorum and $\zeta$ Puppis, can produce enough ionizing photons to sustain the observed size of the IVS-Gum nebula. The exact amount of momentum injection depends on the details of the HII region models and the radiation trapping factor $f_{\rm trap}$, as discussed in \ref{subsubsec:HII region}, a detailed investigation is reserved for future study. We also quantitatively analyzed the role of supernova feedback and estimated that $1.7_{-1.1}^{+2.2}$ supernova events would be required if supernova is the primary driver of the IVS expansion.

\item Assuming the shell is primarily driven by supernovae, we derived a dynamical age of $1.6_{-0.6}^{+1.5}$ Myr. We examined potential sources of supernova feedback: the companion of the runaway star $\zeta$ Puppis, the pulsar RX J0720.4-3125, and two sets of prominent young star clusters. Tracing these sources back in time, we found that the companion of $\zeta$ Puppis likely exploded as a supernova outside the current shell boundary, making it an unlikely energy source for the IVS. The pulsar RX J0720.4-3125 is unlikely to be a contributor either, as it did not pass close to the shell center. However, the two sets of young clusters could have produced 1.16 or 1.21 supernova events in the past 3 Myr \citep[private communication]{Swiggum2024}, making them a likely source of supernova feedback powering the expansion of the IVS.
\end{itemize}

This study presents a detailed analysis of the stellar feedback mechanisms shaping the IVS, along with constraints on its evolutionary history. Our findings offer new insights into the dynamic environment of both the IVS and the Gum Nebula. This work can also serve as a prototype for exploring other regions of the ISM characterized by complex stellar feedback.

\begin{acknowledgments}
We thank all the people who have made this paper possible. We thank Emily Hunt for insightful discussions on the recognition of nearby stellar clusters. We appreciate the constructive feedback from the referee John Bally that have improved this paper. Annie Gao and T. K. Sridharan acknowledge support from the NASA ADAP grant 80NSSC23K0779 titled ``The Galactic Life Cycle: Gum Nebula Cometary Globules and Molecular Clouds." Annie Gao appreciates the input from Ralf Klessen on modeling and theoretical simulations. Special thanks to Veome Kapil for invaluable discussions on the project and assistance with technical programming challenges. 

\end{acknowledgments}

\vspace{5mm}

\appendix

\section{Gaussian Smoothing Kernel $\sigma_{\rm smooth}$ choice and uncertainties}\label{appendix: kernel}

\begin{figure}[htb!]
    \includegraphics[width=1\textwidth]{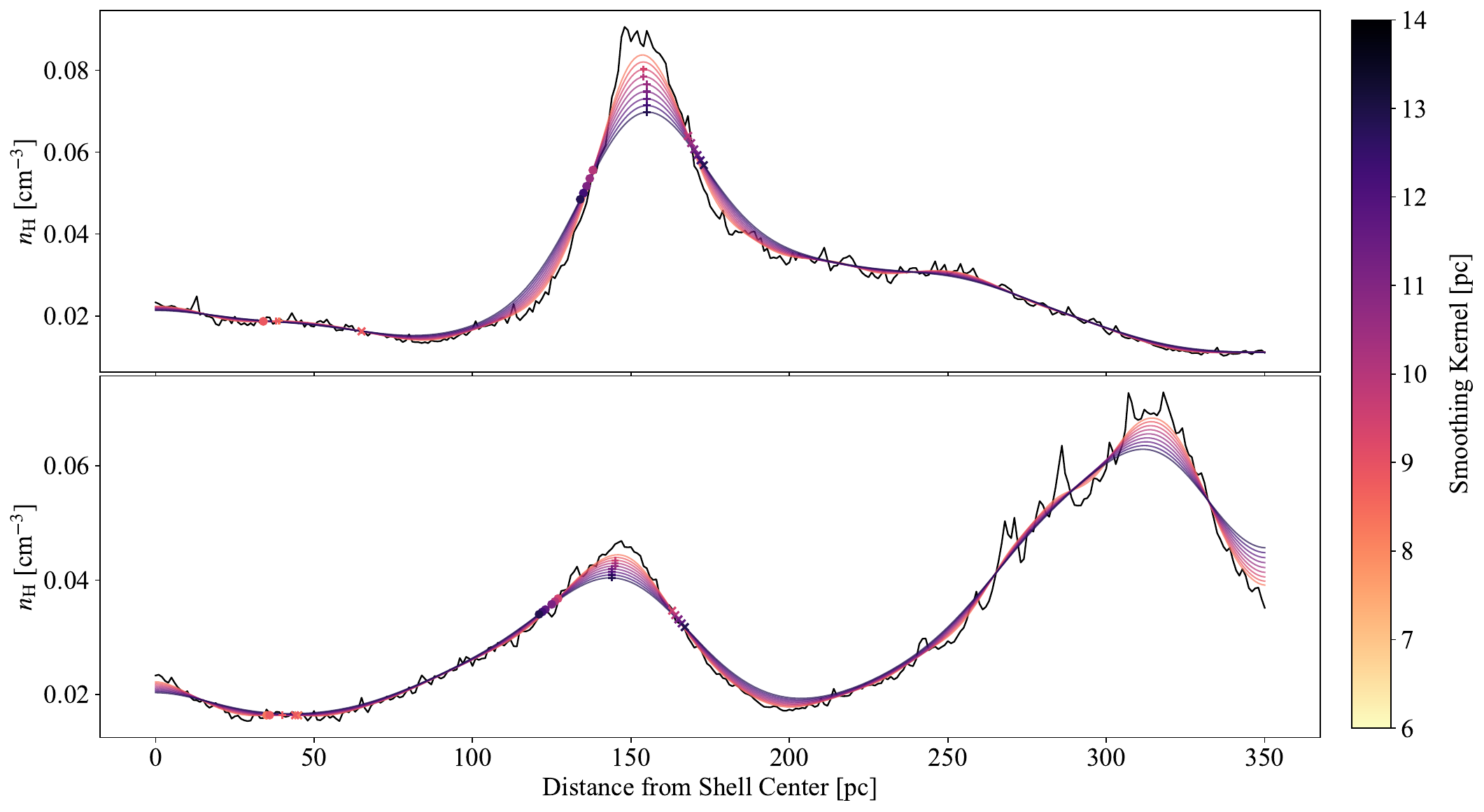}
    \caption{Two examples of $n_{\rm{H}}$ along the ray from IVS center. We show the original profile along with the results smoothed with different Gaussian kernels.}
    \label{fig:smoothing_kernel}
\end{figure}

As mentioned in \S\ref{subsec: boundary}, we apply a Gaussian smoothing kernel to the queried $n_{\rm H}$ to minimize the effects of noise. The smoothing kernel values are empirically determined to mitigate noise while preserving the intrinsic variation of $n_{\rm H}$ along each ray. We vary the standard deviation of the smoothing kernel between 6 pc and 14 pc to the mean 3D dust map from E24 account for systematic uncertainty, and apply the fiducial smoothing kernel of 10 pc to different 3D dust map samples from E24 to estimate the statistical uncertainty. We find sampling from the dust map has a relatively small effect on the boundary definition compared to the choice of smoothing kernel, differing by an order of magnitude. In Figure \ref{fig:smoothing_kernel}, we present two sample density profiles for which we applied our boundary-finding algorithm using different smoothing kernels. The plot shows that when the kernel is small, the smoothed density profile is largely affected by noise, leading to unphysical substructures that our algorithm misinterprets as shell geometry. This effect is seen in the profile smoothed with $\sigma_{\rm smooth}=6$~pc (boundaries shown in orange). For kernels of 10 pc and above, the identified shell structures are qualitatively similar (for those smoothed with $\sigma_{\rm smooth}=10$ pc and 14 pc), though the exact positions of the inner and outer boundaries vary slightly with kernel size. The peaks between the inner and outer boundaries are also highlighted in Fig.~\ref{fig:smoothing_kernel}. While the inflection points depend on the chosen smoothing kernel $\sigma_{\rm smooth}$, the local maximum tends to remain stable across $\sigma_{\rm smooth}>6$~pc.

\section{Boundary-finding Method Comparison} \label{appendix: method-compare}
\begin{figure}[htb!]
    \includegraphics[width=1\textwidth]{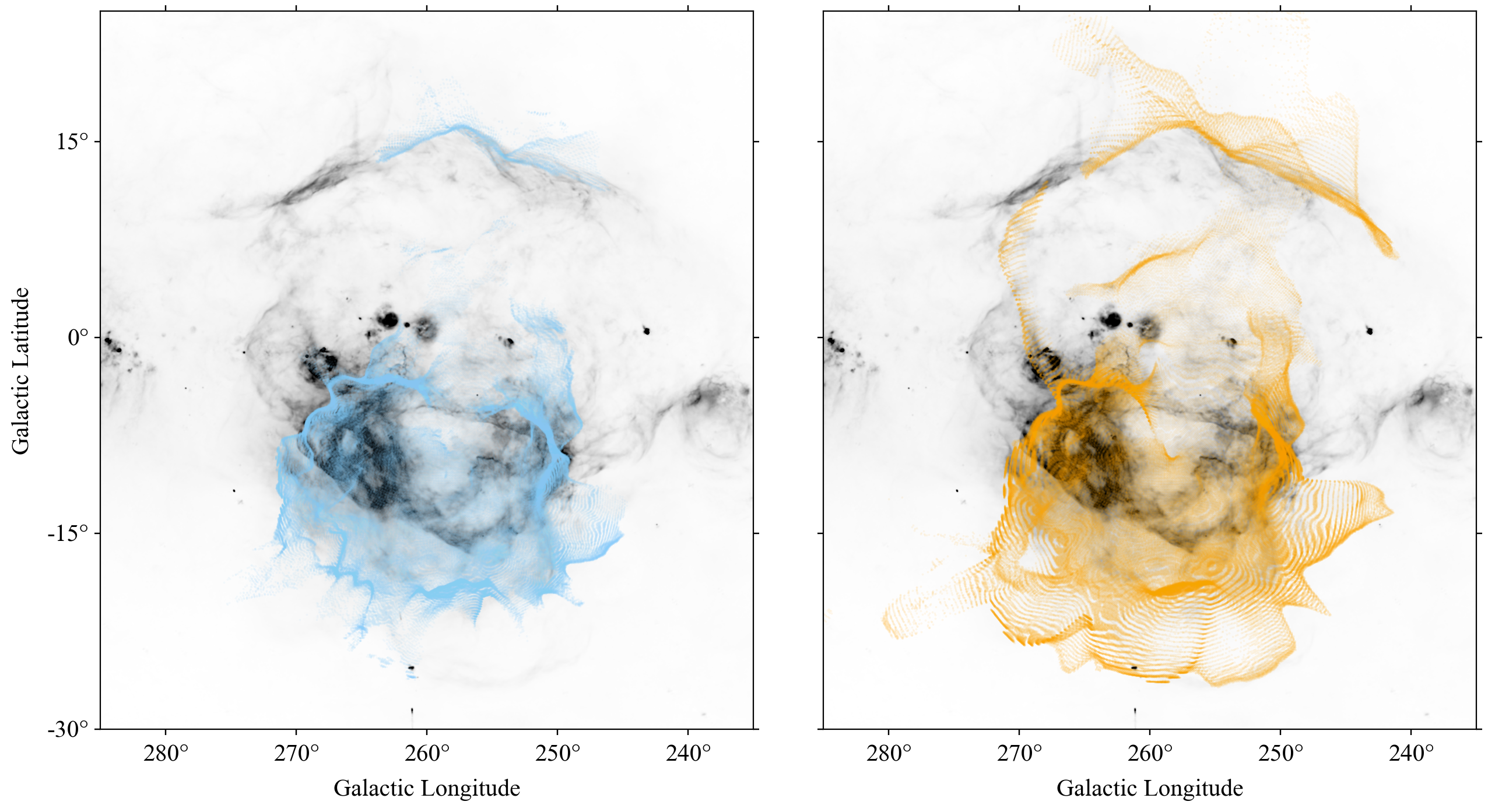}
    \caption{Comparison of projected IVS 3D models. One the left we show the IVS model used in this work (blue) alongside the model constructed with the peak-finding method from \citet{O'Neill2024} (gold), both overlaid on an H$\alpha$ emission maps \citep{Finkbeiner2003}. Both models find similar geometries for the dense bowl-like structure in the plane and the $\rm H\alpha$ filament of the IVS similarly, while differing in their treatment of diffuse regions at around $b \sim -25^\circ$. }
    \label{fig:models_compare}
\end{figure}

Here we compare our IVS model, reconstructed using the inflection-point-finding method outlined in \S\ref{subsec: boundary}, with the peak-finding method described by \citet{O'Neill2024}. The peak-finding approach uses the Python function {\tt find$\textunderscore$peaks} from the {\tt scipy} package, with customized choices of Gaussian smoothing kernel and minimum peak prominence. While the Gaussian smoothing kernel serves a similar function in both methods, the peak prominence parameter in \citet{O'Neill2024} helps preserving high latitude, low-density peaks. Compared to the IVS peak radius $r_{\rm{peak}} = 70_{-22}^{+84}$ pc (95\% C.I.) in this work, the fiducial model constrained by the peak-finding method yields a radius of $r_{\rm{peak}} = 76_{-24}^{+132}$ pc (95\% C.I.). These two approaches provide qualitatively similar median peak radii. Figure \ref{fig:models_compare} shows the resulting models for the peak shell radius overlaid on the H$\alpha$ map from \citet{Finkbeiner2003}. The plot reveals that both models similarly constrain the dense bowl-like region of the nebula between $b=-20^\circ$ and $b=0^\circ$. Both models can identify the $\rm H\alpha$ filament near $b\sim15^\circ$. The primary difference is that the \citet{O'Neill2024} model traces more diffuse features around $b=-25^\circ$. Given that the dense bowl-like section of the nebula contains most of the mass, these model differences should not significantly affect the derived physical properties, such as the mass, expansion momentum, and expansion energy.

\bibliography{sample631}{}
\bibliographystyle{aasjournal}


\end{CJK*}
\end{document}